\documentclass[aps,prl,twocolumn,superscriptaddress,longbibliography]{revtex4-2}
\usepackage{amsmath,amsthm,verbatim,amssymb,amsfonts,amscd,graphicx}
\usepackage[colorlinks=true,citecolor=blue,linkcolor=magenta,urlcolor=blue]{hyperref}
\usepackage{color}
\usepackage[normalem]{ulem}

\newcommand{\blue}[0]{\color{blue}}

\newcommand{\ket}[1]{\ensuremath{\left| #1 \right\rangle}}
\newcommand{\bra}[1]{\ensuremath{\left\langle #1 \right|}}

\newcommand{\bea}{\begin{equation}\begin{aligned}}
\newcommand{\eea}{\end{aligned}\end{equation}}

\usepackage{bm}
\usepackage{placeins}

\begin{document}
\title{Local Impurity Induced Growth and Scrambling in Clean Free Fermions}
\author{Qucheng Gao}
\email{gaoqc@bc.edu}
\affiliation{Department of Physics, Boston College, Chestnut Hill, Massachusetts 02467, USA}
\author{Vikram Ravindranath}
\affiliation{Department of Physics, Boston College, Chestnut Hill, Massachusetts 02467, USA}
\author{Xiao Chen}
\email{chenaad@bc.edu}
\affiliation{Department of Physics, Boston College, Chestnut Hill, Massachusetts 02467, USA}

\begin{abstract}
We study impurity-induced particle growth and scrambling in clean one-dimensional free-fermion systems. We show that a single local impurity can act as a branching source: particle or operator weight propagates coherently into the free bulk, returns to the impurity, and is locally converted into additional degrees of freedom. We develop this branching picture in three complementary settings: a monitored free-fermion model with feedback, a fully unitary interacting particle model, and Heisenberg operator dynamics with an interacting impurity. In the monitored model, we find a feedback-driven transition for both boundary and bulk impurities. In the unitary particle and operator models, a boundary impurity gives rise to an analogous transition from saturation to sustained growth and scrambling. These results reveal how a single impurity can generate complex many-body dynamics in an otherwise clean and free quantum system.

\end{abstract}

\maketitle

\emph{\blue Introduction.--}
Quantum impurity models describe a bath of otherwise free particles coupled locally to a small interacting subsystem. Traditionally, studies of impurity models have focused primarily on how this local coupling modifies the ground-state properties and low-energy phases of the original free system. A paradigmatic example is the Kondo effect, in which a localized magnetic impurity qualitatively modifies the low-energy properties of a metal and, consequently, its low-temperature transport behavior~\cite{kondo1964resistance,hewson1997kondo}. More recently, quantum dynamics has attracted considerable attention, particularly from the perspective of quantum information~\cite{amico2008entanglement,eisert2015quantum,swingle2018unscrambling,fisher2023random}. Inspired by these developments, it is natural to ask: can a localized impurity or perturbation induce nontrivial dynamics in an otherwise simple free-fermion system?

This question was recently explored in free-fermion systems with spatial and temporal randomness~\cite{gao2024information,gao2024scrambling}. In generic interacting quantum many-body systems, a simple local operator typically evolves under Heisenberg time evolution into a highly nonlocal and complex operator, a process known as information scrambling~\cite{hayden2007black,sekino2008fast,shenker2014black,roberts2015localized,shenker2015stringy,maldacena2016bound,hosur2016chaos,roberts2018operator,nahum2017quantum,nahum2018operator,von2018operator}. By contrast, in free-fermion systems, the dynamics is governed by noninteracting quasiparticles, and a local operator remains simple apart from ballistic spreading~\cite{calabrese2005evolution}. Remarkably, Refs.~\cite{gao2024information,gao2024scrambling} showed that in one and two dimensions, a single local interaction can nevertheless generate nontrivial scrambling of a local operator in otherwise free random dynamics, albeit through a slow diffusive process. In dimensions $d\geq 3$, tuning the impurity strength was further shown to induce an escape-to-scrambling transition~\cite{gao2024scrambling}. These phenomena admit an effective description in terms of a classical random walk with a source term.

An important open question is whether analogous impurity-induced scrambling can occur in \emph{clean} systems. In this setting, the bulk fermion operator undergoes coherent quantum-walk dynamics rather than a classical random walk. The competition between ballistic propagation away from the impurity and local growth upon return is therefore qualitatively different, and it is not obvious whether a single impurity can generate sustained complexity.

To address this question and build a transparent physical picture, we first introduce a monitored free-fermion model in which measurements followed by feedback act only on a finite number of sites. This minimal setting captures the key competition between coherent propagation in the free bulk and local growth near the impurity. Using large-scale simulations, we find a dynamical phase transition as the monitoring rate is varied. We then turn to a fully unitary interacting particle model, where the same mechanism survives in a genuine Hamiltonian setting and gives rise to richer dynamical behavior. Finally, guided by these two models, we study Heisenberg operator dynamics in a generic impurity Hamiltonian, which provides the natural framework for describing impurity-induced operator growth in a free-fermion bulk.

Across these three settings, we identify a common mechanism controlled by coherent bulk propagation and local impurity-induced growth. Their competition can drive a transition even in one dimension and produces phenomena richer than those found in random free-fermion dynamics. Taken together, our results provide a unified picture of how a single impurity can generate complexity in clean free-fermion systems.

\begin{figure}
    \centering
    \includegraphics[width=0.99\linewidth]{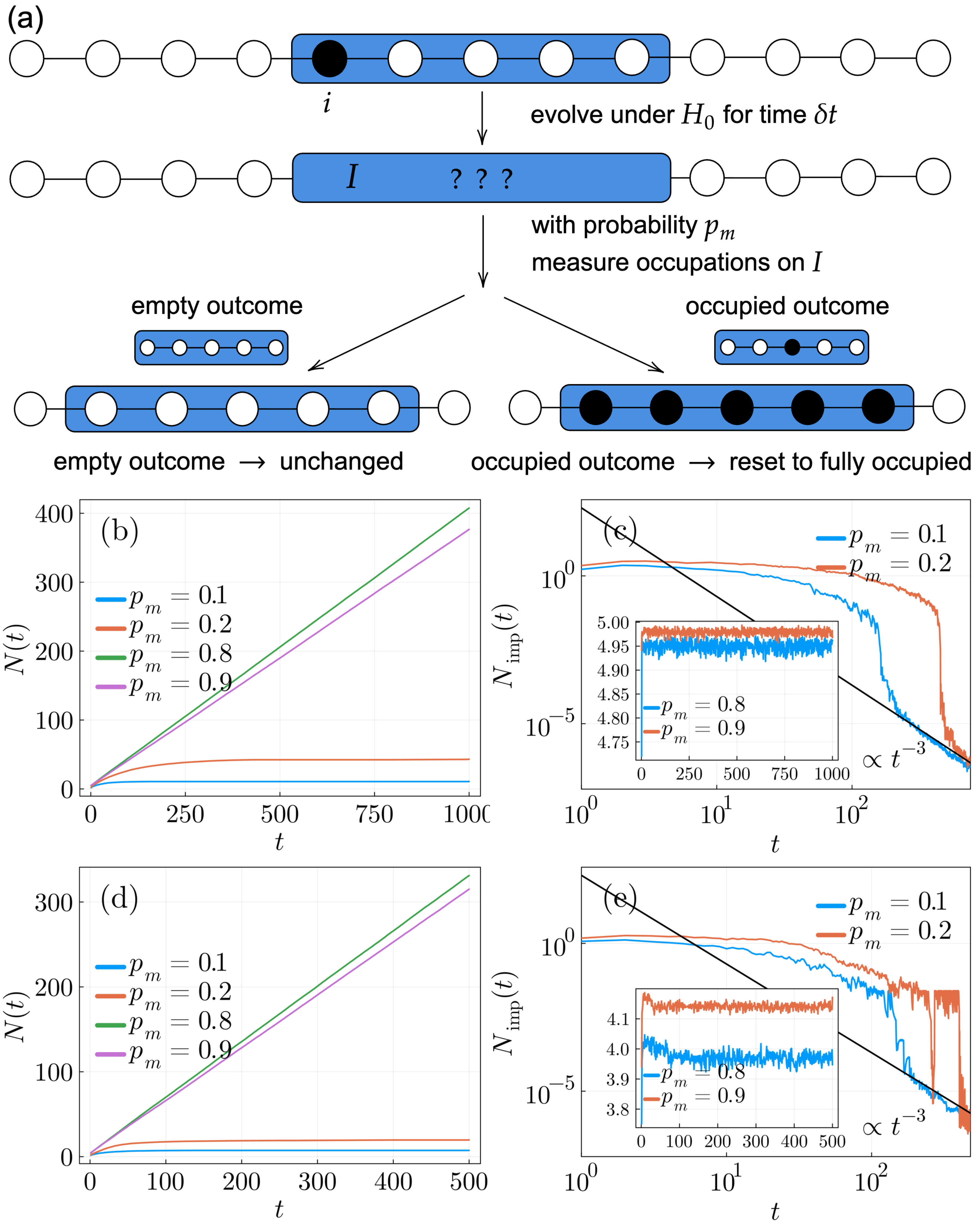}
    \caption{
    Monitored model with feedback.
    (a) Schematic illustration of the protocol.
    At each time step, the system first evolves unitarily for a time interval $\delta t=0.5$, after which measurements are performed at impurity sites with probability $p_m$.
    (b),(c) Boundary impurity: total particle number $N(t)$ and impurity occupation $N_{\mathrm{imp}}(t)$.
    (d),(e) Bulk impurity: total particle number $N(t)$ and impurity occupation $N_{\mathrm{imp}}(t)$.
    In (c) and (e), the main panels are shown on log-log scales for weak monitoring, while the insets show large $p_m$ on linear scales. The black solid lines indicate the power-law decay $t^{-3}$. Here, $L=1000$, and we average over $200$ samples.
    }
    \label{fig:monitored}
\end{figure}

\emph{\blue Monitored model with feedback.--}
We first introduce a monitored free-fermion model with local feedback. This model provides a minimal setting in which the two ingredients of the impurity problem can be cleanly separated: coherent propagation in the free bulk and local particle production at the impurity.

We consider a one-dimensional chain of spinless fermions with open boundary conditions. The impurity region is a finite cluster of $m$ consecutive sites,
\bea
    I=\{i,i+1,\dots,i+m-1\}.
\eea
In the numerics below we take $m=5$. Between feedback steps, the system evolves under the nearest-neighbor hopping Hamiltonian
\bea\label{eq:hopping_H0}
H_0=\sum_{j=1}^{L-1}\left(c_j^\dagger c_{j+1} +c_{j+1}^\dagger c_j\right),
\eea
where $c_j^\dagger$ and $c_j$ create and annihilate a fermion on site $j$.

The dynamics is defined in discrete time steps of duration $\delta t$. During each step, the system first evolves unitarily under $H_0$ for a time $\delta t$. After this free-evolution step, the impurity region is monitored with probability $p_m$. When monitoring occurs, we measure the occupation number on each site in the region $I$. If all sites in $I$ are empty, no reset is applied. If at least one site in $I$ is occupied, the impurity cluster is reset to the fully occupied configuration, as illustrated in Fig.~\ref{fig:monitored}(a). Thus a successful detection event converts any nonzero occupation in the impurity region into a fully occupied cluster localized at the impurity.

Starting from a single particle initially placed in the impurity region, we study the total particle number
\bea
    N(t)=\sum_{j=1}^{L} N_j(t),
\eea
and the impurity occupation
\bea
    N_{\mathrm{imp}}(t)=\sum_{j\in I} N_j(t),
\eea
where $N_j(t)$ denotes the occupation on site $j$ at time $t$. The corresponding state evolution can be simulated efficiently because the bulk evolution is quadratic, while the measurement and feedback act only on the finite impurity cluster~\cite{bravyi2004lagrangian,ravindranath2025free}.

The total particle number $N(t)$ provides a direct diagnostic of whether the feedback can sustain long-time growth. Representative results are shown in Figs.~\ref{fig:monitored}(b) and \ref{fig:monitored}(d) for boundary and bulk impurities, respectively. For weak monitoring, $N(t)$ initially increases but eventually saturates. For sufficiently large $p_m$, by contrast, $N(t)$ continues to grow over the accessible time window and is approximately linear at late times. This behavior indicates a transition between a saturating regime and an active regime with sustained particle production.

To identify the quantity that controls this transition, we examine the impurity occupation $N_{\mathrm{imp}}(t)$, shown in Figs.~\ref{fig:monitored}(c) and \ref{fig:monitored}(e). In the weak-monitoring regime, we find
\bea
    N_{\mathrm{imp}}(t)\sim t^{-3}
\eea
at long times for both boundary and bulk impurities. For sufficiently large $p_m$, $N_{\mathrm{imp}}(t)$ no longer decays to zero within the accessible time window, but instead approaches a nonzero value. These results suggest that the long-time dynamics is controlled by the return of particle weight to the impurity region: in the saturating regime the returning weight eventually vanishes, while in the active regime the impurity remains persistently occupied and continues to seed new outgoing particles. We now formulate this mechanism more explicitly.

\emph{\blue Branching picture.--}
The monitored dynamics admits a simple branching interpretation. Particles created near the impurity propagate through the free bulk under $H_0$; when part of the wave packet returns and is detected, feedback repopulates the impurity cluster and emits a new packet of particles. Thus the impurity acts as a local branching source, while the free bulk controls the probability of return.

This picture contains two ingredients. The first is the return kernel: the probability that a particle emitted from the impurity is found again in the impurity region after time $t$. We denote this probability by $P(t)$. The second is the local branching rule: once weight returns to the impurity, the local dynamics at the impurity can convert it into several particles near the impurity. The long-time behavior is determined by the competition between return and branching. If the total returning weight is too small, or if the local branching is too weak, the process dies out and $N(t)$ saturates. If the branching is strong enough, the impurity becomes self-sustaining and $N(t)$ continues to grow.

For the underlying quantum walk in one dimension, the return probability has the long-time power-law form
\bea
    P(t)\sim t^{-\alpha}.
\eea
For a quantum walker initially placed at the boundary, $\alpha=3$, while for one initially placed in the bulk, $\alpha=1$; see the Supplemental Material for the derivation. This distinction is important because the integrated return weight
\bea
    \int^\infty dt\, P(t)
\eea
is finite for $\alpha>1$ but divergent for $\alpha\leq 1$.
The interpretation of this integrated return weight in terms of recurrence probability and the branching criterion is discussed in the Supplemental Material.
When $\alpha>1$, a particle emitted from the impurity has only a finite total amount of future return weight. Weak local branching therefore cannot automatically sustain growth, and a genuine threshold is possible: below the threshold the branching process is too weak to sustain growth and $N(t)$ saturates, whereas above the threshold the impurity repeatedly reseeds the bulk and $N(t)$ grows.

By contrast, when $\alpha\leq 1$, the integrated return weight diverges. The bulk case, with $P(t)\sim t^{-1}$, is therefore marginal. In this case, the simple threshold argument based on a finite total return weight does not apply in the same way. This distinction will be important below when we compare monitored and fully unitary dynamics.


In the monitored model, repeated measurements modify the effective return kernel. Monitoring interrupts the coherent local return amplitude and produces an effective renewal process, giving
\bea
    P(t)\sim t^{-3}
\eea
for both boundary and bulk impurities; see the Supplemental Material. This accounts for the $t^{-3}$ decay of $N_{\mathrm{imp}}(t)$ in Figs.~\ref{fig:monitored}(c) and \ref{fig:monitored}(e). Since this kernel is integrable, varying $p_m$ can drive the monitored feedback transition.

\begin{figure}
    \centering
    \includegraphics[width=0.99\linewidth]{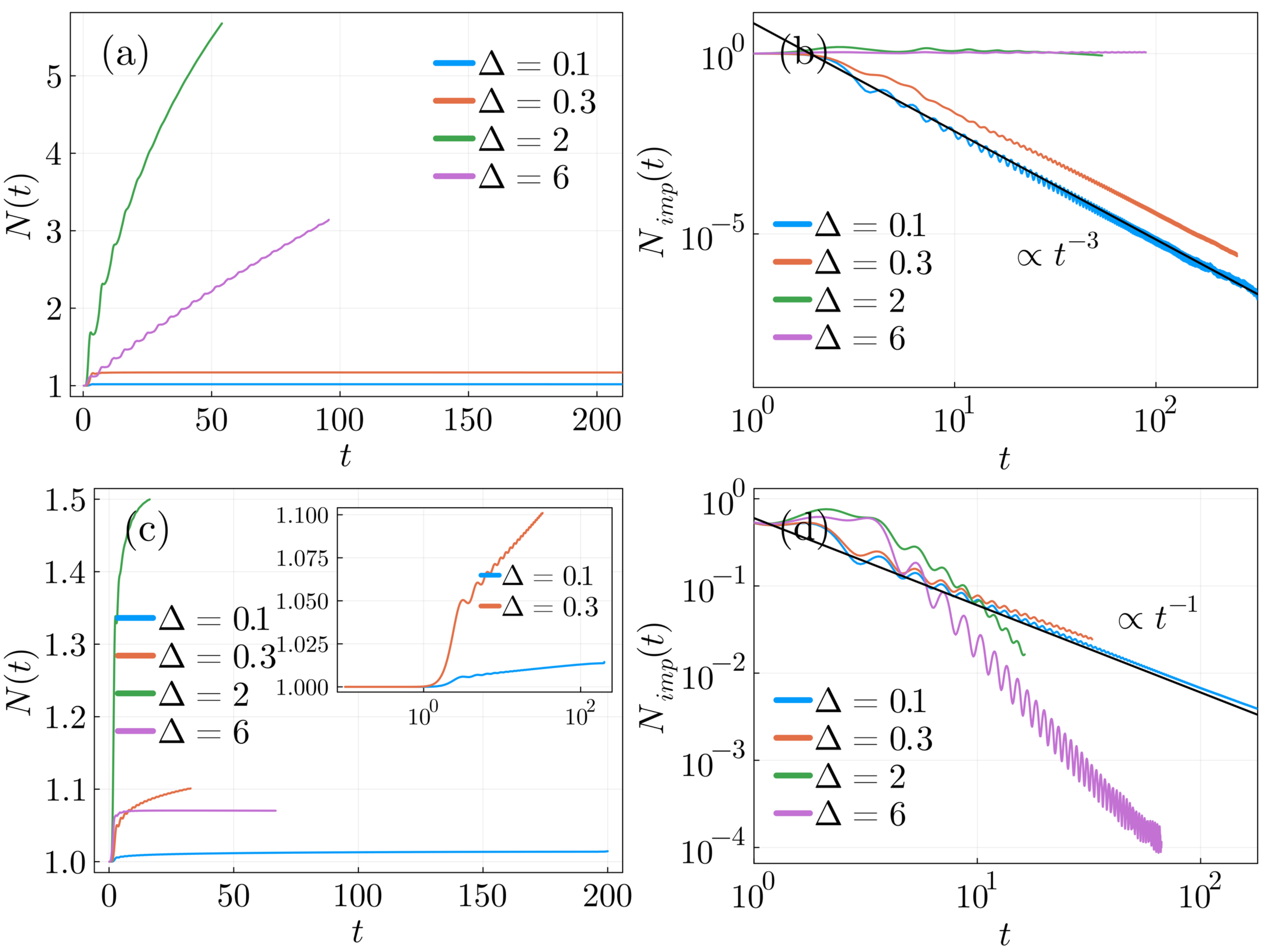}
    \caption{
    Interacting particle model.
    (a),(b) Boundary impurity: total particle number $N(t)$ and impurity occupation $N_{\mathrm{imp}}(t)$.
    (c),(d) Bulk impurity: total particle number $N(t)$ and impurity occupation $N_{\mathrm{imp}}(t)$.
    Results are shown for $L=400$.
    In (b) and (d), the black solid lines indicate the power-law decays $t^{-3}$ and $t^{-1}$, respectively.
    The inset in (c) shows the small-$\Delta$ data on a linear-log scale, highlighting the continued slow growth of $N(t)$ over the accessible time window.
    }
    \label{fig:particle_dynamics}
\end{figure}

\emph{\blue Interacting particle model.--}
We now examine the same branching mechanism in a fully unitary setting, where particle production is generated coherently by a local interaction rather than by an explicit measurement-and-feedback rule that resets the local state. In this case, the impurity coherently converts one particle into multiple particles; in the model below, one particle is converted into three. The resulting growth is controlled by the interplay between this local interaction and the return probability of the free bulk.

We consider
\bea
    H = H_0 + H_{\mathrm{imp}},
\eea
where $H_0$ is the free hopping Hamiltonian in Eq.~\eqref{eq:hopping_H0}, and
\bea
H_{\mathrm{imp}} = \Delta\left( c_i^\dagger c_{i+1}^\dagger c_{i+2}^\dagger c_{i+3} + c_{i+3}^\dagger c_{i+2}c_{i+1}c_i \right).
\eea
The impurity term locally changes the particle number by two while preserving fermion parity. Starting from a single particle initially placed at the impurity,
\bea
    |\Psi(0)\rangle = c_i^\dagger |0\rangle,
\eea
we study the total particle number $N(t)$, the impurity occupation $N_{\mathrm{imp}}(t)$, 

The particle-production current $J(t)$ satisfies
\bea
    \frac{dN(t)}{dt}=2J(t),
\eea
and is discussed in the Supplemental Material.
These quantities are computed using matrix-product-state (MPS) simulations implemented with ITensor~\cite{itensor,itensor-r0.3}.



We first consider a boundary impurity. In this case the free return probability obeys $P(t)\sim t^{-3}$. Since this return law is integrable, the branching picture predicts that weak impurity interactions cannot sustain indefinite growth. This is consistent with Figs.~\ref{fig:particle_dynamics}(a) and \ref{fig:particle_dynamics}(b): for small $\Delta$, $N_{\mathrm{imp}}(t)\sim t^{-3}$ and $N(t)$ saturates at long times. The current $J(t)$ exhibits the same power-law decay, as shown in the Supplemental Material. 
As $\Delta$ is increased, the dynamics deviates from this weak-coupling behavior and $N(t)$ shows stronger growth. This stronger growth is accompanied by rapid entanglement growth, which limits the accessible time window of the MPS simulation.

We next consider a bulk impurity. Here the return probability is instead 
$P(t)\sim t^{-1}$. 
This return law is marginal, since its long-time integral diverges logarithmically. Repeated returns can therefore accumulate over time, and even weak impurity interactions can in principle produce continued, although slow, particle growth.


As shown in Figs.~\ref{fig:particle_dynamics}(c) and \ref{fig:particle_dynamics}(d), $N(t)$ continues to increase over the accessible time window, with the inset of Fig.~\ref{fig:particle_dynamics}(c) highlighting the slow growth at small $\Delta$. In the same regime, $N_{\mathrm{imp}}(t)\sim t^{-1}$, and the current $J(t)$ follows the same scaling. At longer times, the logarithmically divergent return weight suggests that the system may cross over to a more strongly growing regime, although this regime is difficult to access numerically.

This comparison highlights both the similarities and differences between monitored and unitary impurity dynamics. In the monitored model, both boundary and bulk cases are governed by an effective integrable return law, $P(t)\sim t^{-3}$, and hence both can exhibit a phase transition as the feedback strength is varied. In the unitary model, the boundary impurity is still controlled by the integrable return law $P(t)\sim t^{-3}$, while the bulk impurity retains the marginal return law $P(t)\sim t^{-1}$. The bulk unitary case therefore does not exhibit the same simple saturation-to-growth transition. Instead, it is marginal: slow growth is already visible at weak impurity strength at early times, and no finite threshold is expected within this branching picture. A complementary diagnostic is provided by the full particle-number distribution, which shows sparse branching for boundary impurities and slow drift toward larger particle number in the marginal bulk case; see the Supplemental Material.

\begin{figure}
    \centering
    \includegraphics[width=0.99\linewidth]{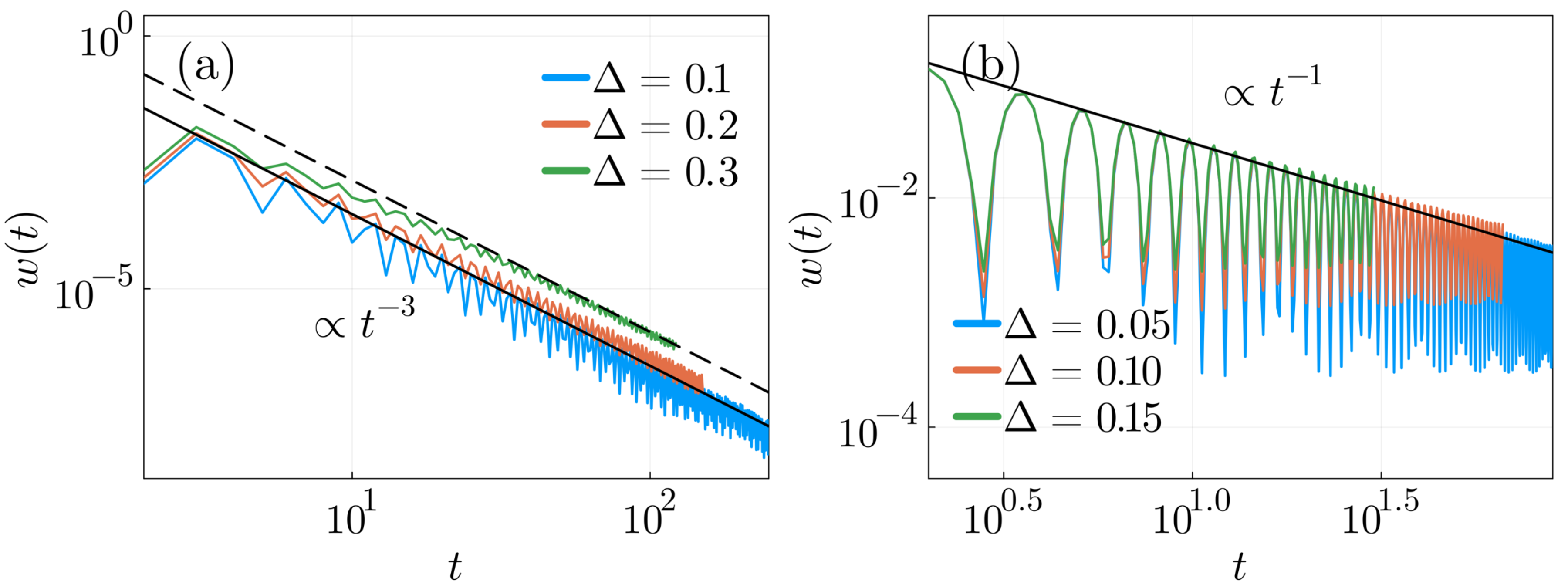}
    \caption{
    Impurity-sensitive operator weight $w(t)$ in the operator model with $H=H_0+\Delta N_i N_{i+1}$.
    (a) Boundary impurity, with $i=1$ and $O(0)=N_1$, for $\Delta=0.1,0.2,0.3$ and $L=400$.
    (b) Bulk impurity, with $i=L/2$ and $O(0)=N_{L/2}$, for $\Delta=0.05,0.10,0.15$ and $L=200$.
    Solid and dashed lines indicate the power-law decays $w(t)\sim t^{-3}$ in (a) and $w(t)\sim t^{-1}$ in (b).
    }
    \label{fig:op}
\end{figure}

\emph{\blue Operator dynamics.--}
We finally turn to Heisenberg operator dynamics. Expanding $O(t)$ in a Majorana-string basis defines an operator size, which is conserved under the free Hamiltonian but can grow in the presence of interactions. Impurity-induced scrambling therefore corresponds to a sustained transfer of operator weight from short strings to longer strings. The analogue of $N_{\rm imp}(t)$ is the impurity-sensitive operator weight $w(t)$, which measures the portion of the operator currently supported near the impurity and hence available for further local branching. Its precise definition is given in the Supplemental Material. In the free limit, $w(t)$ is proportional to the local return probability at long times.

We consider
\bea
    H = H_0 + H_{\mathrm{imp}},
\eea
where $H_0$ is the free hopping Hamiltonian in Eq.~\eqref{eq:hopping_H0}, and
\bea
    H_{\mathrm{imp}} = \Delta N_i N_{i+1}.
\eea
Starting from a simple local operator at the impurity,
\bea
    O(t)=e^{iHt}O(0)e^{-iHt},
    \qquad
    O(0)=N_i,
\eea
we represent $O(t)$ as a matrix-product operator (MPO) and simulate its time evolution using ITensor~\cite{itensor,itensor-r0.3}. We compute $w(t)$ for both boundary and bulk impurities. In the weak-impurity regime, the branching picture predicts that $w(t)$ is controlled by the same return law that governs the particle-number dynamics. As shown in Figs.~\ref{fig:op}(a) and \ref{fig:op}(b), we find
\bea
    w(t)\sim t^{-\alpha},
\eea
with $\alpha=3$ at the boundary and $\alpha=1$ in the bulk. 
Thus, for a boundary impurity, the integrable return law keeps impurity-induced operator-growth events sparse. By contrast, for a bulk impurity, the return law is marginal. The observed $w(t)\sim t^{-1}$ scaling should therefore be interpreted as a weak-coupling behavior over the accessible time window, rather than definitive evidence for an asymptotic decay law.

\begin{figure}
    \centering
    \includegraphics[width=0.99\linewidth]{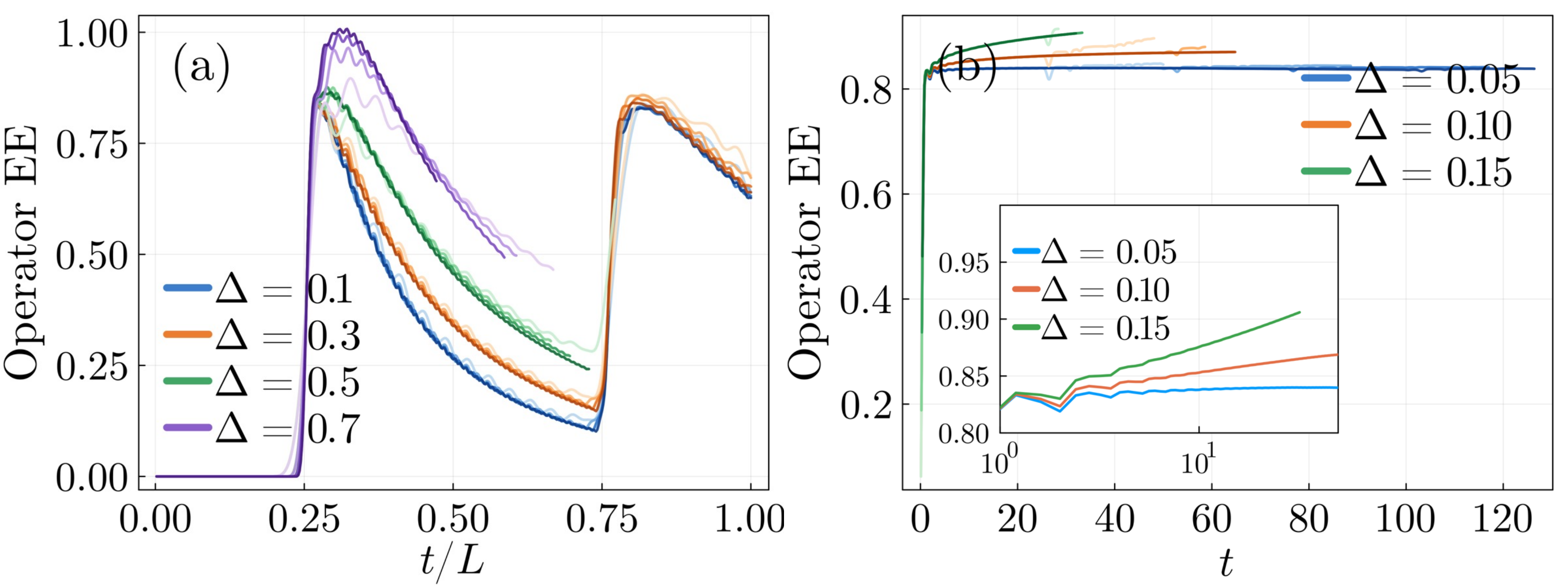}
    \caption{
    Half-system operator entanglement entropy in the operator model.
    (a) Boundary impurity.
    (b) Bulk impurity.
    For small $\Delta$, the entropy grows slowly.
    Inset: $L=200$ data on a linear-log scale, highlighting the slow late-time growth.
    From light to dark colors: $L=50,100,150,200$.
    The late-time upturn in (a) near $t/L\simeq0.75$ is a finite-size reflection effect from the opposite boundary.
    }
    \label{fig:op_EE}
\end{figure}

Directly extracting the operator size from the MPO simulation is numerically challenging. We therefore use the half-system operator entanglement entropy as an indirect diagnostic of impurity-induced complexity. For a boundary impurity, we find that the operator entanglement entropy saturates to a small value that is essentially independent of system size, as shown in Fig.~\ref{fig:op_EE}(a). This behavior is consistent with sparse branching events controlled by the integrable boundary return law. By contrast, when the impurity is placed in the bulk, the operator entanglement entropy grows slowly at early times, with a form close to logarithmic growth; see Fig.~\ref{fig:op_EE}(b). This is consistent with the configuration-entropy calculation presented in the Supplemental Material.

Finally, as the impurity strength $\Delta$ is increased, the operator entanglement entropy grows much more rapidly in time, even at early times, for both boundary and bulk impurities. This enhancement of operator entanglement provides further evidence that the local interaction acts as a branching source for operator growth, in agreement with the branching picture developed above.




\emph{\blue Discussion.--}
We have developed a branching picture for impurity-induced dynamics in clean free-fermion systems. Across the monitored model, the interacting particle model, and Heisenberg operator dynamics, the long-time behavior is controlled by the return probability to the impurity and the local branching rule generated by the impurity. This framework explains how a single local perturbation can produce saturation, sustained growth, or slow marginal growth in an otherwise free system.

In one dimension, the location of the impurity plays a crucial role. For the unitary dynamics studied in this work, a boundary impurity has an integrable return law, allowing a genuine transition between a saturation regime and a growth/scrambling regime as the impurity strength is increased. By contrast, a bulk impurity has a marginal return law. As a result, weak local branching already accumulates over long times, and the sharp saturation-to-growth transition found at the boundary is absent. Instead, the bulk impurity produces slow marginal growth at weak coupling and crosses over to faster growth as the impurity strength is increased.

The same reasoning can be generalized to higher dimensions. For a local impurity in a clean free-fermion system, the coherent return probability decays more rapidly with time as the spatial dimension increases. Consequently, for $d\geq 2$, the return kernel is integrable, and a finite-threshold impurity-induced growth transition is expected both for boundary and bulk impurities. Details of the higher-dimensional return-probability analysis are presented in the Supplemental Material.

The branching picture is not expected to be completely universal. We have checked that it is robust for several choices of local impurity interactions, as well as in a Floquet realization, as discussed in the Supplemental Material. At the same time, the picture can break down when the impurity cannot be described as a local branching source coupled to a well-defined return kernel. Examples include parity-breaking impurity terms that generate nonlocal Jordan--Wigner strings and systems with topological zero modes at the boundary. These cases are also discussed in the Supplemental Material. Nevertheless, within its regime of applicability, the branching framework provides a simple physical mechanism by which a single impurity can induce particle growth and operator scrambling in an otherwise free-fermion system.

\vspace{5pt}
    \emph{\blue Acknowledgements.} 
We thank Eliot Heinrich, Serge Florens, Hanmeng Zhan, and Thomas Scaffidi for insightful discussions. We gratefully acknowledge computing resources from Research Services at Boston College and the assistance provided by Wei Qiu.
\bibliography{manuscript.bbl}

\clearpage
\onecolumngrid
\begin{center}
{\large \textbf{Supplemental Material}}
\end{center}
\vspace{0.5cm}
\FloatBarrier

\pagenumbering{arabic} 

\renewcommand{\thepage}{S\arabic{page}}
\renewcommand{\thesection}{S\arabic{section}}
\renewcommand{\theequation}{S\arabic{equation}}
\renewcommand{\thefigure}{S\arabic{figure}}
\renewcommand{\thetable}{S\arabic{table}}

\setcounter{equation}{0}
\setcounter{figure}{0}
\setcounter{table}{0}

\appendix
\section{Derivation of the free-fermion exponents $\alpha=1,3$}
\label{sec:SM_alpha}

In this section we derive the power-law exponents $\alpha$ that control the long-time decay of the free return probability $P(t)$ near the impurity. This same free return law underlies the decay of the impurity occupation $N_{\mathrm{imp}}(t)$ discussed in the main text.

For the hopping Hamiltonian
\bea
    H_0=\sum_{j=1}^{L-1}\left(c_j^\dagger c_{j+1}+c_{j+1}^\dagger c_j\right),
\eea
the single-particle dynamics is governed by the $L\times L$ hopping matrix $h$ defined by
\bea
    H_0=\sum_{j,k} h_{jk} c_j^\dagger c_k.
\eea
We denote the single-particle basis by $\ket{j}\equiv c_j^\dagger\ket{0}$, and define the single-particle propagator
\bea\label{eq::def_single_U}
    U_{j\ell}(t)\equiv \bra{j}e^{-ih t}\ket{\ell}.
\eea

Consider a particle initially created at the impurity site $i$,
\bea
    |\psi(0)\rangle=c_i^\dagger|0\rangle .
\eea
The occupation at site $j$ and time $t$ is then
\bea
    N_j(t)=|\bra{j}e^{-iht}\ket{i}|^2 = |U_{ji}(t)|^2.
\eea

For an impurity region $I$ consisting of a finite number of sites, we define the free return probability
\bea
    P_I(t)=\sum_{j\in I} N_j(t)=\sum_{j\in I}|U_{ji}(t)|^2.
\eea
Since the impurity region contains only a finite number of sites, its long-time decay exponent is the same as that of any fixed site near the impurity. It therefore suffices to derive the scaling using the return probability to the initial site,
\bea\label{eq::definition_S}
    P_i(t)\equiv |U_{ii}(t)|^2,
\eea
which we simply denote by $P(t)$ below.

\paragraph{Bulk site (translation-invariant limit).}
For a site far from boundaries, one may use the infinite-chain limit, where
\bea
U_{j\ell}(t)=\frac{1}{2\pi}\int_{-\pi}^{\pi}dk\,e^{ik(j-\ell)-i2t\cos k} = i^{\,j-\ell} J_{j-\ell}(2t),
\eea
with $J_n$ the Bessel function of the first kind. In particular,
\bea
U_{ii}(t)=J_0(2t)\sim \sqrt{\frac{1}{\pi t}} \cos\!\left(2t-\frac{\pi}{4}\right), \qquad (t\to\infty),
\eea
so that, up to bounded oscillatory factors,
\bea
    P(t)=|U_{ii}(t)|^2 \sim t^{-1}.
\eea
This gives the bulk exponent $\alpha=1$.

\paragraph{Boundary site.}
For an open chain, the boundary return amplitude at $i=1$ can be written as
\bea\label{eq:U11_integral}
    U_{11}(t)=\frac{2}{\pi}\int_0^\pi dk\,\sin^2 k\, e^{-i2t\cos k}.
\eea
At long times this integral is dominated by the stationary points at $k=0$ and $k=\pi$. Near $k=0$, $\sin^2 k\simeq k^2$ and $\cos k\simeq 1-\frac{k^2}{2}$, yielding
\bea
    U_{11}(t)\sim e^{-i2t}\int_0^\infty dk\,k^2 e^{it k^2}\propto t^{-3/2}.
\eea
The same scaling arises from the neighborhood of $k=\pi$. Therefore, up to an overall oscillatory factor,
\bea
    U_{11}(t)\sim t^{-3/2},
    \qquad
P(t)=|U_{11}(t)|^2\sim t^{-3}, \qquad (t\to\infty),
\eea
which gives the boundary exponent $\alpha=3$.

\paragraph{Higher dimensions.}
The same stationary-phase argument extends directly to higher dimensions.
For a bulk impurity in a $d$-dimensional clean free-fermion system, the local
return amplitude is a $d$-dimensional momentum integral. Near a generic
nondegenerate stationary point, each momentum direction contributes a factor
$t^{-1/2}$, giving
\bea
    U_{ii}(t)\sim t^{-d/2},\qquad P(t)=|U_{ii}(t)|^2\sim t^{-d}.
\eea
For an impurity on a flat open boundary, the momentum component normal to the
boundary carries the same boundary factor as in Eq.~\eqref{eq:U11_integral},
which changes its contribution from $t^{-1/2}$ to $t^{-3/2}$. The remaining
$d-1$ directions contribute the usual bulk factors. Thus
\bea
    U_{ii}(t)\sim t^{-(d+2)/2},\qquad P(t)\sim t^{-(d+2)} .
\eea
Therefore, for $d\geq 2$, both the bulk and boundary return probabilities are
integrable in time.
In the branching picture, this implies that repeated returns to a local impurity have a finite total weight. Sustained impurity-induced growth is therefore expected only beyond a finite local branching strength, leading to a finite-threshold growth transition.

\section{Integrated return weight and the branching picture}
\label{sec:SM_branching_picture}

In this section we clarify the physical meaning of the integrated return weight that appears in the branching picture.
The key distinction in the main text is whether the return probability $P(t)$ has a finite or divergent time integral.
This distinction is the quantum analogue of the distinction between transient and recurrent return in a classical random walk.

Consider first a classical random-walk picture.
Let $p_r$ denote the recurrence probability, namely the probability that a particle returns to the impurity region at least once after leaving it.
If a return occurs, the particle can leave again and attempt to return again.
The expected total number of returns is then
\bea
    \mathcal N_{\rm ret}
    =
    p_r+p_r^2+p_r^3+\cdots
    =
    \frac{p_r}{1-p_r}.
\eea
Equivalently,
\bea
    p_r=\frac{\mathcal N_{\rm ret}}{1+\mathcal N_{\rm ret}}.
\eea
Thus a finite expected number of returns corresponds to $p_r<1$, meaning that there is a nonzero escape probability.
By contrast, a divergent expected number of returns corresponds to $p_r=1$, meaning that returns are recurrent and continue to accumulate over time.

This relation gives a simple interpretation of the criterion $n_{\rm br}p_r=1$ used in our previous work on impurity-induced scrambling in random free-fermion dynamics~\cite{gao2024scrambling}.
If each return to the impurity produces, on average, $n_{\rm br}$ outgoing branches, then the effective reproduction factor is
\bea
    n_{\rm br}p_r .
\eea
The transition occurs when
\bea
    n_{\rm br}p_r=1 .
\eea
For $n_{\rm br}p_r<1$, branching is too weak to sustain long-time growth.
For $n_{\rm br}p_r>1$, repeated returns and branching events can sustain long-time growth.

In the present clean quantum setting, this classical relation should be viewed only as a heuristic analogy: the return probability $P(t)$ comes from coherent quantum propagation, and the impurity dynamics is not literally a classical branching process. Nevertheless, the integrated return weight
\bea
    \mathcal N_{\rm ret}^{(q)}
    \sim
    \int^\infty dt\, P(t)
\eea
plays the same physical role as the expected number of returns. If $\mathcal N_{\rm ret}^{(q)}$ is finite, an emitted wave packet has only a finite total amount of future return weight near the impurity. Equivalently, there is an effective escape probability, so local branching must be strong enough to sustain growth. This gives a finite-threshold transition between saturation and sustained growth.

If instead $\mathcal N_{\rm ret}^{(q)}$ diverges, returns accumulate over time and the finite-threshold argument no longer applies. The one-dimensional bulk case, with $P(t)\sim t^{-1}$, is marginal in this sense: the accumulated return weight grows only logarithmically, so weak branching can produce slow growth over accessible times rather than a sharp saturation-to-growth threshold.

Thus the integrability of the return probability is the central criterion in the branching picture. The boundary case $P(t)\sim t^{-3}$ has finite integrated return weight and can support a finite threshold, whereas the bulk case $P(t)\sim t^{-1}$ is marginal and leads to slow accumulated growth at weak impurity strength.

\section{Why the monitored model gives $N_{\mathrm{imp}}(t)\sim t^{-3}$ for both boundary and bulk impurities}
\label{sec:SM_monitored_alpha}

In this section we provide a heuristic argument for why, in the monitored dynamics, the impurity occupation decays as
\bea
    N_{\mathrm{imp}}(t)\sim t^{-3},
\eea
for both boundary and bulk impurities. The key point is that monitoring acts as repeated resets of the local amplitude, leading to a renewal description of the long-time return near the impurity.

\paragraph{Free return law.}
Let $P_0(t)$ denote the return probability in the impurity region in the absence of monitoring. For a one-dimensional free-fermion system, the long-time behavior is
\bea
    P_0(t)\sim t^{-\alpha},
\eea
with
\bea
    \alpha=
    \begin{cases}
1, & \text{bulk},\\ 3, & \text{boundary}.
    \end{cases}
\eea
Equivalently, in terms of the return amplitude $A_0(t)$,
\bea
    A_0(t)\sim t^{-\alpha/2}.
\eea

\paragraph{Effect of monitoring.}
Suppose that the impurity is monitored with a nonzero rate $p_m>0$. Each monitoring event destroys the coherent amplitude accumulated in the impurity region. Averaging over monitoring histories, the effective local amplitude obeys a renewal-type equation,
\bea
    A(t)\approx A_0(t)-p_m\sum_{\tau<t} A_0(t-\tau)A(\tau).
\eea
Here $A_0(t)$ is the free return amplitude, and the second term accounts for histories interrupted by a measurement event at some earlier time $\tau$. In Laplace space this becomes
\bea
    \widetilde A(s)\approx \frac{\widetilde A_0(s)}{1+p_m \widetilde A_0(s)}.
    \label{eq:renewal_laplace}
\eea
The long-time behavior is determined by the small-$s$ structure of $\widetilde A_0(s)$.

\paragraph{Bulk impurity.}
For a bulk impurity in one dimension,
\bea
    A_0(t)\sim t^{-1/2},
\eea
so its Laplace transform is singular at small $s$,
\bea
\widetilde A_0(s)\sim c\, s^{-1/2}, \qquad s\to 0,
\eea
with some nonuniversal constant $c$. Substituting this into Eq.~\eqref{eq:renewal_laplace}, we obtain
\bea
    \widetilde A(s)\sim \frac{c\, s^{-1/2}}{1+p_m c\, s^{-1/2}}.
\eea
At sufficiently small $s$, the denominator is dominated by the second term, and therefore
\bea
    \widetilde A(s)\sim \frac{1}{p_m}-\frac{1}{p_m^2 c}s^{1/2}+\cdots.
\eea
The constant term contributes only at short times, while the nonanalytic $s^{1/2}$ term determines the long-time tail. Inverting the Laplace transform gives
\bea
    A(t)\sim t^{-3/2}.
\eea
Therefore the survival probability behaves as
\bea
    P(t)=|A(t)|^2\sim t^{-3}.
\eea

Thus, although the free bulk return probability is only $P_0(t)\sim t^{-1}$, any nonzero monitoring rate changes the asymptotic behavior to
\bea
    P(t)\sim t^{-3}.
\eea

\paragraph{Boundary impurity.}
For a boundary impurity in one dimension,
\bea
    A_0(t)\sim t^{-3/2},
\eea
so the small-$s$ expansion of the Laplace transform is regular,
\bea
    \widetilde A_0(s)=a_0-a_1 s^{1/2}+\cdots,
\eea
with nonuniversal constants $a_0$ and $a_1$. Substituting into Eq.~\eqref{eq:renewal_laplace}, we find
\bea
    \widetilde A(s)=\frac{a_0-a_1 s^{1/2}+\cdots}{1+p_m(a_0-a_1 s^{1/2}+\cdots)}.
\eea
Since the denominator remains finite at $s\to 0$, the leading nonanalytic term is still proportional to $s^{1/2}$. Therefore the long-time behavior remains
\bea
    A(t)\sim t^{-3/2},
\eea
and hence
\bea
    P(t)\sim t^{-3}.
\eea
In other words, monitoring changes only the prefactor of the boundary survival law, but not the exponent.

\paragraph{Impurity occupation.}
Since the impurity occupation is determined by the return probability in the impurity region, it has the same long-time behavior,
\bea
    N_{\mathrm{imp}}(t)\sim P(t)\sim t^{-3}.
\eea
This holds for both boundary and bulk impurities.

\begin{figure}[t]
    \centering
    \includegraphics[width=0.95\linewidth]{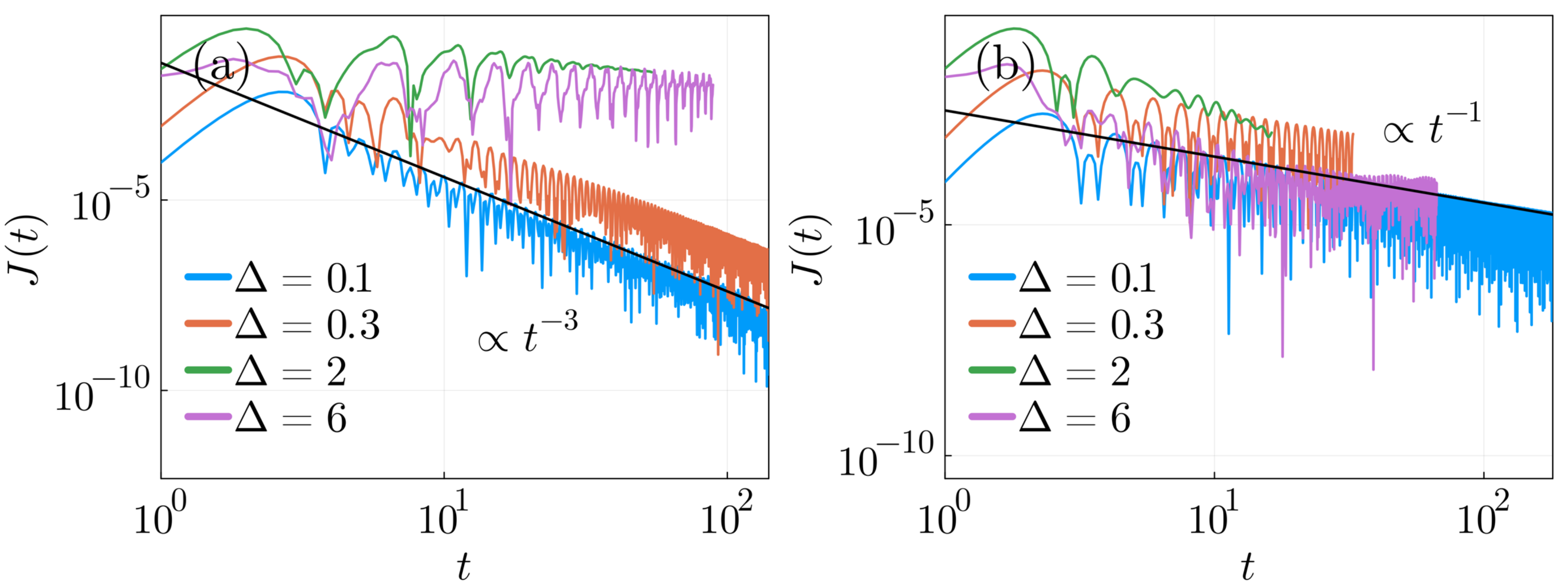}
    \caption{
    Particle-production current $J(t)$ in the interacting particle model for $L=400$.
    (a) Boundary impurity.
    (b) Bulk impurity.
    Black solid lines indicate the power-law decays $t^{-3}$ and $t^{-1}$, respectively.
    }
    \phantomsection\label{sm_fig:current}
\end{figure}

\section{Particle-production current in the interacting particle model}
\label{sec:SM_Jt}
In this section we derive the expression for the particle-production current $J(t)$ and present numerical results for $J(t)$ in the interacting particle model for both boundary and bulk impurities.

We consider the interacting particle model
\bea
    H = H_0 + H_{\mathrm{imp}},
\eea
with
\bea
H_{\mathrm{imp}} = \Delta\left( c_i^\dagger c_{i+1}^\dagger c_{i+2}^\dagger c_{i+3} + c_{i+3}^\dagger c_{i+2} c_{i+1} c_i \right).
\eea
The total particle number operator is
\bea
    N=\sum_j N_j=\sum_j c_j^\dagger c_j.
\eea
Since the bulk Hamiltonian $H_0$ conserves particle number, the time dependence of $\langle N(t)\rangle$ is entirely generated by $H_{\mathrm{imp}}$:
\bea
\frac{d}{dt}\langle N(t)\rangle = i\langle [H,N]\rangle = i\langle [H_{\mathrm{imp}},N]\rangle .
\eea

Using
\bea
[N,c_i^\dagger c_{i+1}^\dagger c_{i+2}^\dagger c_{i+3}] = 2\,c_i^\dagger c_{i+1}^\dagger c_{i+2}^\dagger c_{i+3},
\eea
and
\bea
[N,c_{i+3}^\dagger c_{i+2} c_{i+1} c_i] = -2\,c_{i+3}^\dagger c_{i+2} c_{i+1} c_i,
\eea
we obtain
\bea
\frac{d}{dt}\langle N(t)\rangle = 2i\Delta\Big[ \langle c_{i+3}^\dagger c_{i+2} c_{i+1} c_i\rangle - \langle c_i^\dagger c_{i+1}^\dagger c_{i+2}^\dagger c_{i+3}\rangle \Big].
\eea
This motivates the definition
\bea
    J(t)
    \equiv
i\Delta\Big[ \langle c_{i+3}^\dagger c_{i+2} c_{i+1} c_i\rangle - \langle c_i^\dagger c_{i+1}^\dagger c_{i+2}^\dagger c_{i+3}\rangle \Big],
\eea
for which
\bea
    \frac{d}{dt}\langle N(t)\rangle = 2J(t).
    \label{eq:dNdt_J}
\eea

Representative results are shown in Fig.~\ref{sm_fig:current}. For boundary impurities [Fig.~\ref{sm_fig:current}(a)], we find
\bea
    J(t)\sim t^{-3},
\eea
for small $\Delta$, consistent with the scaling of $N_{\mathrm{imp}}(t)$ discussed in the main text. For bulk impurities [Fig.~\ref{sm_fig:current}(b)], we similarly observe
\bea
    J(t)\sim t^{-1},
\eea
for small $\Delta$ over the accessible time window. As $\Delta$ increases, $J(t)$ deviates from these weak-coupling scalings in both cases.

\section{Numerical evaluation of the particle-number distribution}
\label{sec:SM_particle_distribution}
In this section we describe the numerical procedure used to obtain the fixed-time particle-number distribution in the interacting particle model.

Let
\bea
    N=\sum_{j=1}^{L}N_j
\eea
denote the total particle-number operator. At a fixed time $t$, the probability distribution for finding $n$ particles is
\bea
P_t(n)=\langle \delta_{N,n}\rangle, \qquad n=0,1,\dots,L.
\eea
Rather than projecting directly onto each particle-number sector, it is convenient to compute this distribution from the full counting statistics of $\hat N$.

We introduce the characteristic function
\bea
\chi_t(\theta) = \left\langle e^{i\theta N}\right\rangle = \left\langle \prod_{j=1}^{L} e^{i\theta N_j}\right\rangle .
\eea
One has
\bea
e^{i\theta N_j} = \mathbb I + \left(e^{i\theta}-1\right)N_j.
\eea
Therefore $\chi_t(\theta)$ can be evaluated efficiently in the MPS representation by applying the corresponding onsite phase gate to every site and computing the overlap with the original state.

Because the total particle number takes integer values between $0$ and $L$, the distribution can be reconstructed exactly from a discrete Fourier transform. Choosing
\bea
    M=L+1,
    \qquad
    \theta_k=\frac{2\pi k}{M},
    \qquad
    k=0,1,\dots,L,
\eea
we evaluate
\bea
    \chi_t(\theta_k)=\left\langle e^{i\theta_k N}\right\rangle,
\eea
and then obtain
\bea
P_t(n) =
    \frac{1}{M}
\sum_{k=0}^{L} e^{-i\theta_k n}\,\chi_t(\theta_k), \qquad n=0,1,\dots,L.
\eea
In the numerics, small negative values generated by truncation and floating-point errors are set to zero, after which the distribution is normalized so that
\bea
    \sum_{n=0}^{L} P_t(n)=1.
\eea

\begin{figure}
    \centering
    \includegraphics[width=0.65\linewidth]{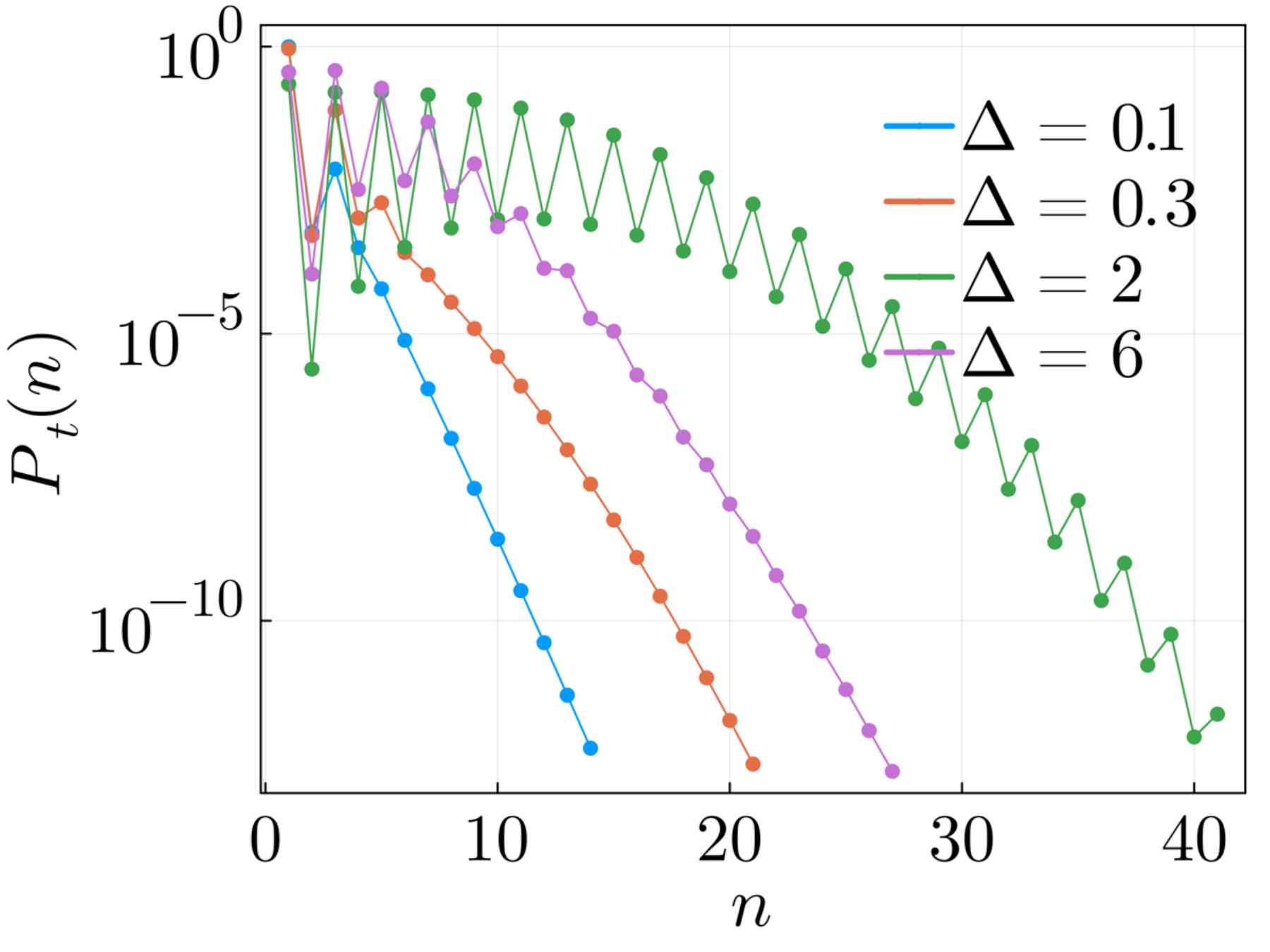}
    \caption{
    Fixed-time particle-number distribution $P_t(n)$ in the interacting
    particle model for a boundary impurity.
    For $\Delta=0.1,0.3,2,6$, the distributions are evaluated at the
    latest numerically accessible times:
    $t=100,100,64,87$, respectively.
    The distributions are consistent with an approximately exponential
    tail. Here, $L=100$.
    }
    \label{sm:fig_particle_dist}
\end{figure}

Representative results for a boundary impurity are shown in Fig.~\ref{sm:fig_particle_dist}. For each value of $\Delta$, we evaluate the distribution at the latest numerically accessible time. In the boundary case, the return law is integrable, so branching events remain sparse in the weak-coupling regime and the total particle number does not drift strongly at late times. Consistent with this picture, the distribution develops an approximately exponential tail. This behavior reflects rare, repeated local branching events near the impurity.

For a bulk impurity, by contrast, the distribution remains visibly time-dependent over the accessible time window and drifts slowly toward larger particle number, as discussed below. This behavior is consistent with the marginal return law $P(t)\sim t^{-1}$, for which repeated returns accumulate logarithmically in time.

\begin{figure}
    \centering
    \includegraphics[width=0.99\linewidth]{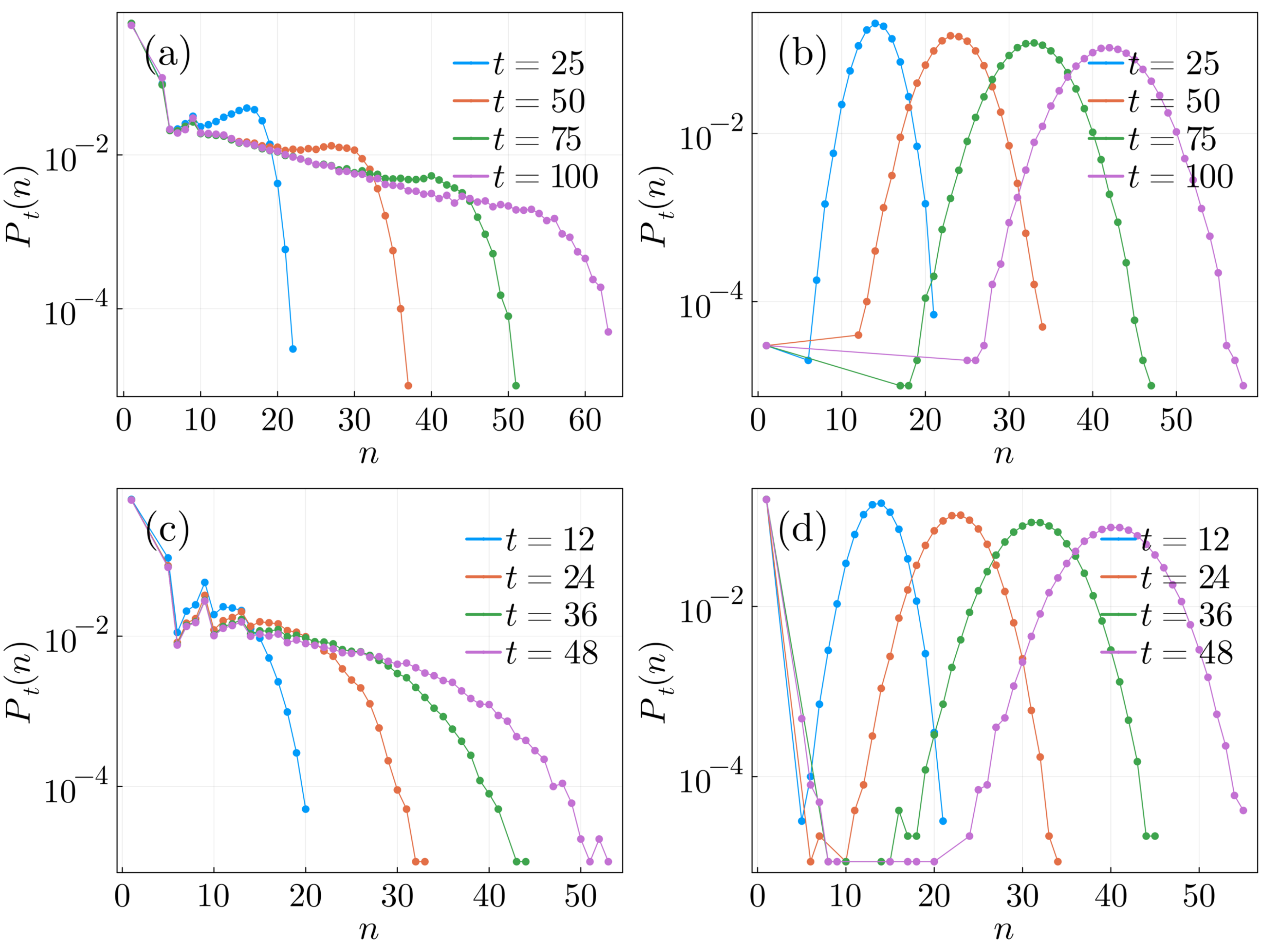}
    \caption{
    Time dependence of the particle-number distribution $P_t(n)$ in the monitored model for $L=100$.
    (a),(b) Boundary impurity with weak monitoring $p_m=0.1$ and strong monitoring $p_m=0.9$, respectively.
    (c),(d) Bulk impurity with weak monitoring $p_m=0.1$ and strong monitoring $p_m=0.9$, respectively.
    For weak monitoring, the distributions at different times nearly collapse, indicating that the particle number distribution has reached a nearly stationary form.
    For strong monitoring, $P_t(n)$ develops a peak at larger $n$, and the peak shifts toward larger $n$ as time increases, reflecting sustained growth.
    Here, we average over $100000$ samples.
    }
    \phantomsection\label{sm_fig:monitored_dist_time}
\end{figure}

\begin{figure}
    \centering
    \includegraphics[width=0.99\linewidth]{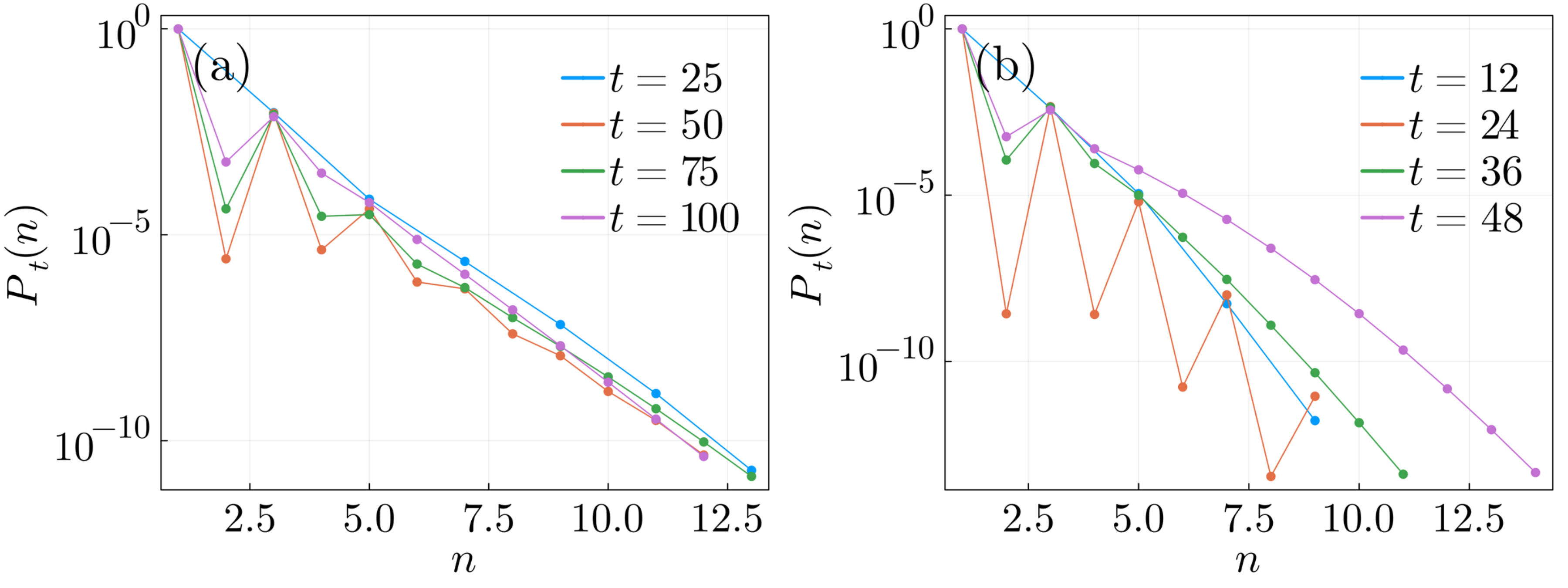}
    \caption{
    Time dependence of the particle-number distribution $P_t(n)$ in the interacting particle model for $L=100$ and $\Delta=0.1$.
    (a) Boundary impurity.
    The distributions at different times nearly collapse, consistent with saturation of the total particle number.
    (b) Bulk impurity.
    The distribution gradually shifts toward larger $n$ as time increases, indicating that the particle number continues to grow over the accessible time window.
    }
    \phantomsection\label{sm_fig:particle_dist_time}
\end{figure}

\section{Time dependence of the particle-number distribution}
\label{sec:SM_dist_time}

In this section we examine the time dependence of the particle-number distribution $P_t(n)$. This provides a useful way to distinguish regimes in which the distribution has reached a nearly stationary form from regimes in which the particle number continues to drift to larger values.

We first consider the monitored model. Representative results are shown in Fig.~\ref{sm_fig:monitored_dist_time}. For weak monitoring [Figs.~\ref{sm_fig:monitored_dist_time}(a) and (c)], $p_m=0.1$, the distributions for both boundary and bulk impurities are nearly unchanged over the time window shown. This is consistent with the behavior of $N(t)$ in the main text: the total particle number saturates, and the full distribution approaches a nearly stationary form.

By contrast, for strong monitoring [Figs.~\ref{sm_fig:monitored_dist_time}(b) and (d)], $p_m=0.9$, the distribution does not approach a stationary exponential tail over the same time window. Instead, it develops a peak at finite $n$, with an approximately Gaussian shape around the peak. As time increases, this peak shifts toward larger particle number. This behavior reflects the sustained-growth regime of the monitored dynamics, where the feedback repeatedly repopulates the impurity region and the total particle number continues to increase.

We next consider the interacting particle model at weak coupling, $\Delta=0.1$. The results are shown in Fig.~\ref{sm_fig:particle_dist_time}. For the boundary impurity, the distributions at different times nearly collapse. This is consistent with the integrable return law $P(t)\sim t^{-3}$ and with the saturation of the total particle number observed in the main text.

For the bulk impurity, the behavior is different. The distribution gradually drifts toward larger particle number as time increases. This reflects the marginal return law $P(t)\sim t^{-1}$, for which the integrated return weight grows logarithmically with time. Therefore, unlike the boundary case, the bulk distribution has not reached a stationary form within the accessible time window.

\begin{figure}
    \centering
    \includegraphics[width=0.99\linewidth]{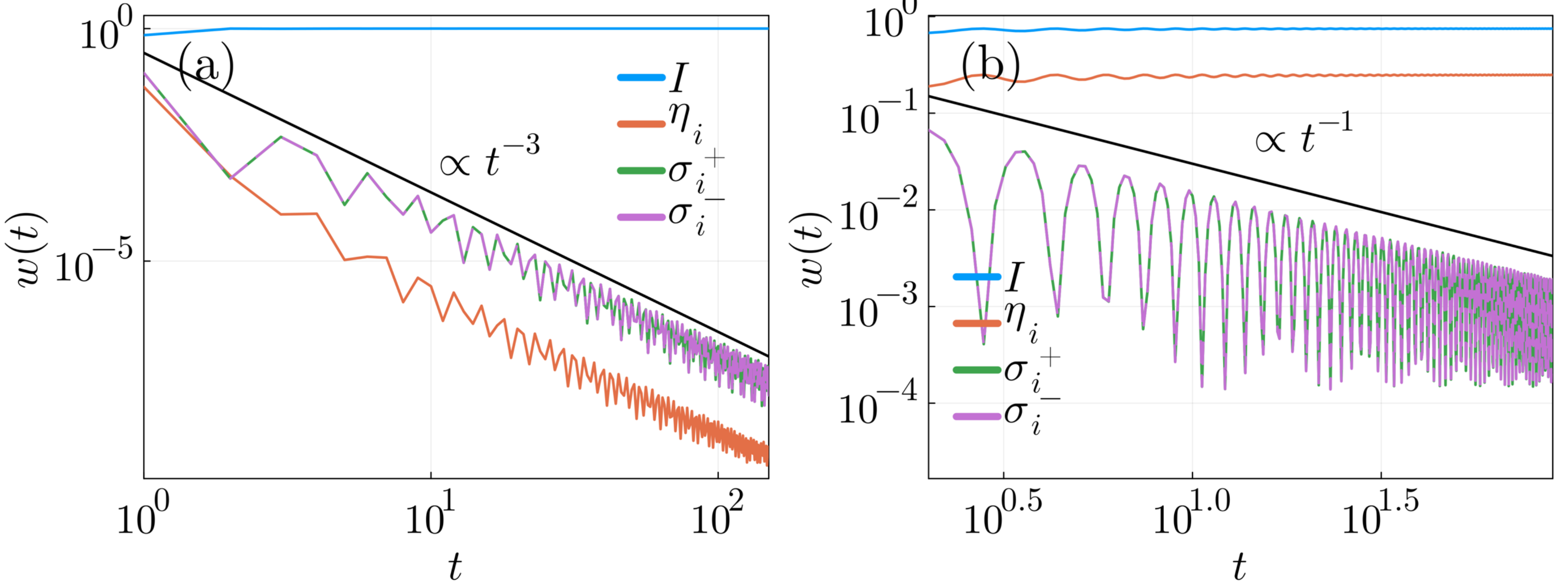}
    \caption{
    Local operator weights in the operator model with $H=H_0+\Delta N_i N_{i+1}$.
    (a) Boundary impurity, with $i=1$, $O(0)=N_1$, $\Delta=0.1$, and $L=400$.
    (b) Bulk impurity, with $i=L/2$, $O(0)=N_{L/2}$, $\Delta=0.05$, and $L=200$.
    The impurity-sensitive weight is defined as $w(t)=w_+(t)+w_-(t)$.
    Black solid lines indicate the power-law decays $t^{-3}$ in (a) and $t^{-1}$ in (b).
    }
    \phantomsection\label{sm_fig:comp}
\end{figure}

\section{Definition of the impurity-sensitive operator weight $w(t)$}
\label{sec:SM_op_components}

In this section we give a precise definition of the operator weight $w(t)$ used in the main text.

We consider the operator dynamics generated by
\bea
    H = H_0 + H_{\mathrm{imp}},
    \qquad
    H_{\mathrm{imp}} = \Delta N_i N_{i+1},
\eea
starting from the initial operator
\bea
    O(0)=N_i.
\eea
Under Heisenberg evolution,
\bea
    O(t)=e^{iHt}O(0)e^{-iHt}.
\eea

To characterize the operator content that remains sensitive to the impurity, we focus on the local operator structure at site $i$. Using the operator-state mapping with the Hilbert--Schmidt inner product, we trace out all sites except site $i$ and obtain the reduced operator density matrix $\rho_i(t)$.

A convenient orthonormal local operator basis on site $i$ is
\bea
\left\{ \frac{\mathbb{I}_i}{\sqrt{2}}, \frac{\eta_i}{\sqrt{2}}, \sigma_i^+, \sigma_i^- \right\},
\eea
where
\bea
    \eta_i = 1-2N_i,
\eea
and, under the Jordan--Wigner mapping,
\bea
    \sigma_i^+ = \eta_1\eta_2\cdots \eta_{i-1}\, c_i^\dagger,
    \qquad
    \sigma_i^- = \eta_1\eta_2\cdots \eta_{i-1}\, c_i.
\eea
We then define the corresponding local weights as
\bea
w_{\mathbb I}(t) &=& \mathrm{Tr}\!\left[ \rho_i(t)\, \Big|\frac{\mathbb I_i}{\sqrt2}\Big\rangle\!\Big\rangle \Big\langle\!\Big\langle \frac{\mathbb I_i}{\sqrt2}\Big| \right],\\ w_{\eta}(t) &=& \mathrm{Tr}\!\left[ \rho_i(t)\, \Big|\frac{\eta_i}{\sqrt2}\Big\rangle\!\Big\rangle \Big\langle\!\Big\langle \frac{\eta_i}{\sqrt2}\Big| \right],\\ w_{+}(t) &=& \mathrm{Tr}\!\left[ \rho_i(t)\, |\sigma_i^+\rangle\!\rangle \langle\!\langle \sigma_i^+| \right],\\ w_{-}(t) &=& \mathrm{Tr}\!\left[ \rho_i(t)\, |\sigma_i^-\rangle\!\rangle \langle\!\langle \sigma_i^-| \right].
\eea
By construction,
\bea
    w_{\mathbb I}(t)+w_{\eta}(t)+w_{+}(t)+w_{-}(t)=1.
\eea

The reason this decomposition is useful is that the impurity interaction commutes with the identity and local parity sectors,
\bea
    [H_{\mathrm{imp}},\mathbb I_i]=0,
    \qquad
    [H_{\mathrm{imp}},\eta_i\eta_{i+1}]=0,
\eea
whereas it does not commute with the ladder sectors,
\bea
    [H_{\mathrm{imp}},\sigma_i^+]\neq 0,
    \qquad
    [H_{\mathrm{imp}},\sigma_i^-]\neq 0.
\eea
We therefore define
\bea\label{eq:def_pt}
    w(t)=w_{+}(t)+w_{-}(t),
\eea
which measures the local operator weight at site $i$ that remains sensitive to the impurity interaction. In the language of the main text, $w(t)$ captures the part of the operator that can still participate in further impurity-induced growth. Representative results for the local weights are shown in Figs.~\ref{sm_fig:comp}(a) and (b) for boundary and bulk impurities, respectively. As shown in Fig.~\ref{sm_fig:comp}, the ladder-sector weights $w_+(t)$ and $w_-(t)$ exhibit the same long-time power-law decay as the impurity-sensitive weight $w(t)$.

For concreteness, we project onto site $i$ only, even though the impurity term acts on both $i$ and $i+1$. Including both impurity sites leads to the same physical picture and the same long-time scaling, while the single-site definition above already suffices to isolate the impurity-sensitive operator content.

\subsection{Relation between $w(t)$ and the free return probability}
\label{sec:SM_alpha_operator}

We now relate the free return probability $P(t)$ derived in Eq.~\eqref{eq::definition_S} to the impurity-sensitive operator weight $w(t)$ defined in Eq.~\eqref{eq:def_pt} for the operator model in the noninteracting limit $\Delta=0$.

For a generic quadratic Hamiltonian
\bea
    H_0=\sum_{j,k} h_{jk} c_j^\dagger c_k,
\eea
the Heisenberg evolution of fermion operators is linear:
\bea
\frac{d}{dt}c_j(t)=i[H_0,c_j(t)] =-i\sum_k h_{jk}c_k(t),
    \qquad
\Rightarrow\quad \vec c(t)=e^{-iht}\vec c.
\eea
Equivalently,
\bea
    c_j(t)=\sum_\ell U_{j\ell}(t)c_\ell,
    \qquad
    c_j^\dagger(t)=\sum_\ell U_{j\ell}^*(t)c_\ell^\dagger,
\eea
with the same single-particle propagator $U(t)=e^{-iht}$ defined as in Eq.~\eqref{eq::def_single_U}.

In the operator model we take the initial operator
\bea
    O(0)=N_i=c_i^\dagger c_i,
\eea
and evolve it under $H_0$:
\bea\label{eq:Ni_expansion}
N_i(t)=c_i^\dagger(t)c_i(t) =\sum_{j,k}U_{ij}^*(t)U_{ik}(t)\,c_j^\dagger c_k.
\eea
To extract the impurity-sensitive weight $w(t)$, we recall that in the local operator decomposition at site $i$, the ladder sectors $\sigma_i^\pm$ are generated by terms with exactly one fermionic operator acting on site $i$. In the expansion above, these are precisely the terms
\bea
    c_i^\dagger c_{k\neq i},
    \qquad
    c_{j\neq i}^\dagger c_i,
\eea
whose amplitudes are proportional to $U_{ii}^*(t)U_{ik}(t)$ and $U_{ij}^*(t)U_{ii}(t)$, respectively.

Using the orthogonality of different operator strings, the total ladder-sector weight scales as
\bea
    w(t)
    \propto
2|U_{ii}(t)|^2\sum_{k\neq i}|U_{ik}(t)|^2 = 2|U_{ii}(t)|^2\bigl(1-|U_{ii}(t)|^2\bigr),
\eea
where we used the unitarity of $U(t)$. At long times, when $|U_{ii}(t)|^2\ll 1$, this reduces to
\bea
    w(t)\propto |U_{ii}(t)|^2.
\eea
Therefore the long-time decay of $w(t)$ is controlled by the same free return law,
\bea
    w(t)\sim P(t)\sim t^{-\alpha},
\eea
with
\bea
    \alpha=
    \begin{cases}
1,& i\ \text{in the bulk},\\ 3,& i\ \text{at the boundary}.
    \end{cases}
\eea

\paragraph{Generality for the operator model.}
Although the above derivation was illustrated using the nearest-neighbor hopping model, the resulting exponents are more general on the operator side. The only ingredients are (i) a bounded single-particle spectrum with well-defined band edges, and (ii) an analytic, even dispersion near the band extrema. Under these conditions, the leading low-energy expansion of the dispersion is quadratic, yielding the same bulk and boundary return exponents. Deviations from this behavior arise only when these assumptions are violated, for example by boundary-localized modes, which are discussed below.

\section{Estimate of the configuration entropy in the operator model}
\label{sec:SM_Sconf}

In this section we explain the estimate of the configuration entropy in the operator model used in the main text.

In the branching picture, operator growth comes from repeated returns to the impurity. Each return can trigger a local growth event, while between returns the operator propagates freely in the bulk. When such growth events remain sufficiently sparse, the fixed-time operator-length distribution is expected to develop an exponential tail,
\bea
    P_t(\ell)\propto e^{-\ell/\xi(t)},
\eea
where $\xi(t)$ is the corresponding characteristic length scale.

The behavior of $\xi(t)$ depends on the return law. For $\alpha>1$, and in particular for the boundary case $\alpha=3$, the integrated return weight is finite. In the weak-coupling regime, this leads to a finite characteristic length,
\bea
    \xi(t)\sim O(1).
\eea
For the bulk case $\alpha=1$, the integrated return weight grows logarithmically with time. In the weak-coupling regime studied numerically, this suggests
\bea
    \xi(t)\sim \log t
\eea
over the accessible time window. We do not make a general claim about the true late-time asymptotic behavior in this case.

To estimate the corresponding configuration entropy, we further assume that within a given length sector $\ell$, the operator weight is approximately spread uniformly over the operator configurations accessible within the ballistic light cone.

Let $\Omega_\ell(t)$ denote the number of operator configurations of length $\ell$ available at time $t$. For a one-dimensional light cone of size $vt$, we estimate
\bea
    \Omega_\ell(t)\sim \binom{vt}{\ell}.
\eea
Under this approximation, the configuration entropy is
\bea
    S_{\mathrm{conf}}(t)
    \approx
-\sum_\ell P_t(\ell)\log P_t(\ell) + \sum_\ell P_t(\ell)\log \Omega_\ell(t).
    \label{eq:Sconf_def}
\eea

We now estimate the two terms separately. For the configurational part, using Stirling's approximation and assuming $\ell\ll vt$, we have
\bea
\log \Omega_\ell(t) = \log \binom{vt}{\ell} = \ell \log(vt)-\log \ell! = \ell\log t-\ell\log \ell+O(\ell).
    \label{eq:logOmega_operator}
\eea
Averaging over the exponential distribution gives
\bea
\sum_\ell P_t(\ell)\log \Omega_\ell(t) = O\!\left(\overline \ell(t)\log t\right) + O\!\left(\overline{\ell\log \ell}\right),
\eea
where
\bea
    \overline \ell(t)=\sum_\ell \ell\,P_t(\ell)\sim \xi(t).
\eea
For an exponential distribution, one also has
\bea
    \overline{\ell\log \ell}=O\!\left(\xi(t)\log \xi(t)\right).
\eea
Therefore
\bea
\sum_\ell P_t(\ell)\log \Omega_\ell(t) = O\!\left(\xi(t)\log t\right) + O\!\left(\xi(t)\log \xi(t)\right).
    \label{eq:config_part}
\eea

The first term in Eq.~\eqref{eq:Sconf_def} is the Shannon entropy of the length distribution,
\bea
    H[P_t(\ell)]
    \equiv
    -\sum_\ell P_t(\ell)\log P_t(\ell).
\eea
For an exponential distribution, this contributes only
\bea
    H[P_t(\ell)]=O(\log \xi(t)),
    \label{eq:length_entropy}
\eea
which is subleading compared with Eq.~\eqref{eq:config_part} whenever $\xi(t)$ is finite or grows slowly.

Combining Eqs.~\eqref{eq:config_part} and \eqref{eq:length_entropy}, we obtain
\bea
S_{\mathrm{conf}}(t) = O\!\left(\xi(t)\log t\right) + O\!\left(\xi(t)\log \xi(t)\right) + O\!\left(\log \xi(t)\right).
\eea
The leading contribution is therefore
\bea
    S_{\mathrm{conf}}(t)\sim \xi(t)\log t.
\eea

This gives
\bea
    S_{\mathrm{conf}}(t)\sim
    \begin{cases}
\log t, & \alpha=3,\\ (\log t)^2, & \alpha=1
    \end{cases}
\eea
within the regime described above.

\begin{figure}
    \centering
    \includegraphics[width=0.99\linewidth]{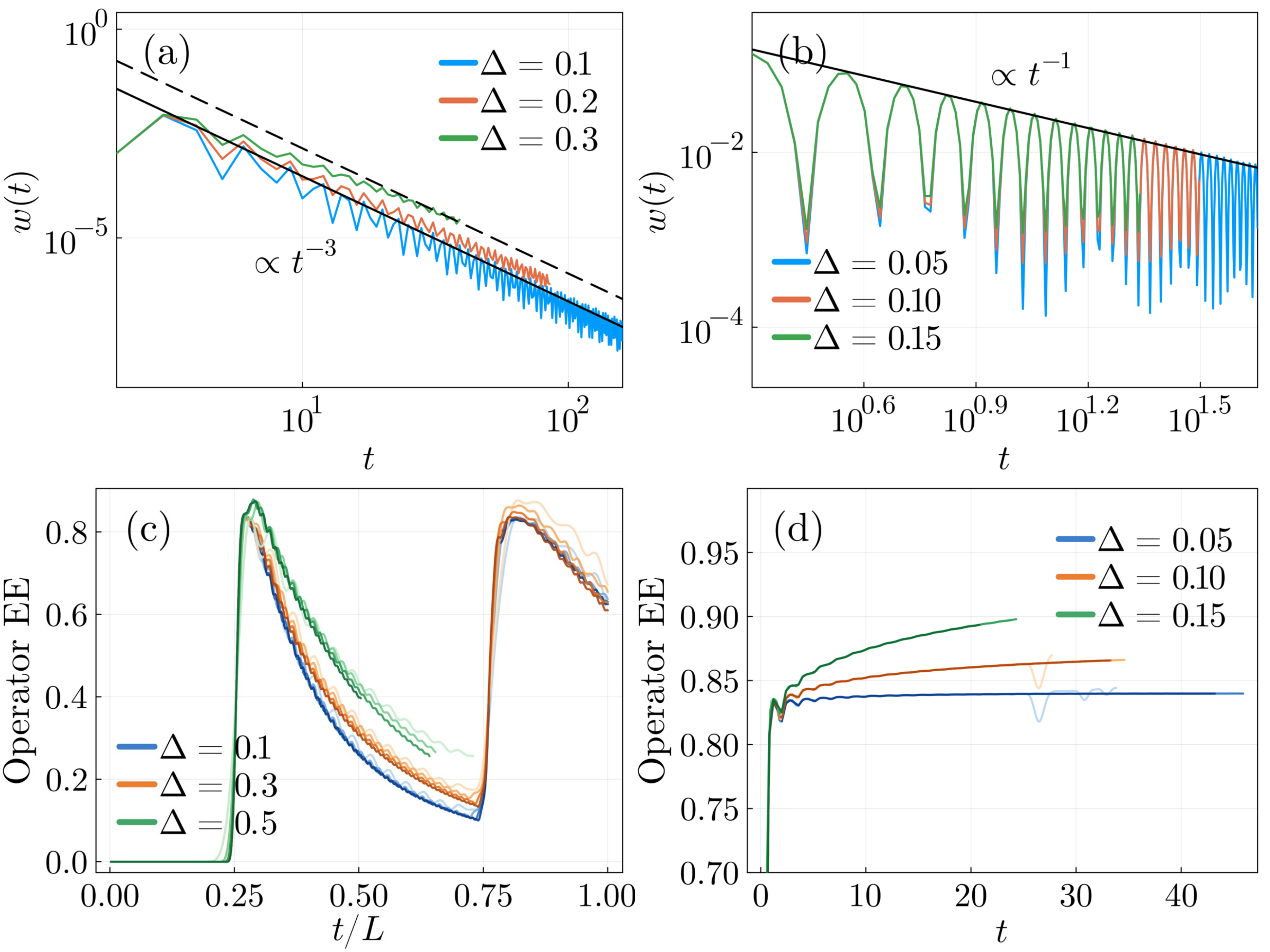}
    \caption{
    Operator dynamics with the impurity interaction $H_{\mathrm{imp}}=\Delta N_i N_{i+1}N_{i+2}$.
    (a),(b) Impurity-sensitive weight $w(t)$ for boundary and bulk impurities, respectively, with $O(0)=N_i$; the system sizes are $L=400$ in (a) and $L=200$ in (b).
    Solid and dashed lines indicate the power-law decays $t^{-3}$ in (a) and $t^{-1}$ in (b).
    (c),(d) Half-system operator entanglement entropy for the same setups.
    From light to dark colors: $L=50,100,150,200$.
    }
    \phantomsection\label{sm_fig:nnn}
\end{figure}

\begin{figure}
    \centering
    \includegraphics[width=0.99\linewidth]{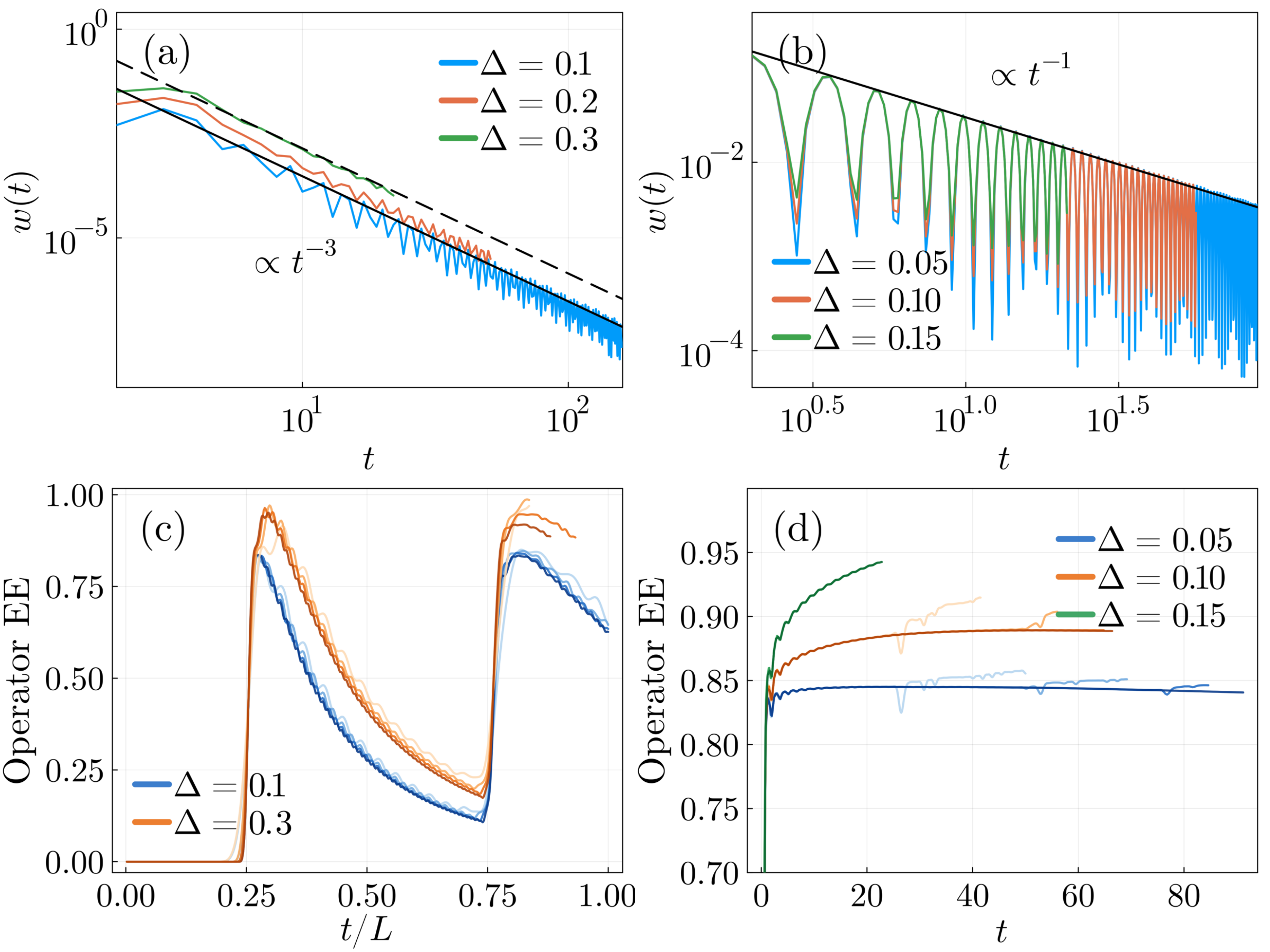}
    \caption{
    Operator dynamics with the impurity interaction
    $H_{\mathrm{imp}}=\Delta(c_i^\dagger c_{i+1}^\dagger c_{i+2}^\dagger c_{i+3}+\mathrm{h.c.})$.
    (a),(b) Impurity-sensitive weight $w(t)$ for boundary and bulk impurities, respectively, with $O(0)=N_i$; the system sizes are $L=400$ in (a) and $L=200$ in (b).
    Solid and dashed lines indicate the power-law decays $t^{-3}$ in (a) and $t^{-1}$ in (b).
    (c),(d) Half-system operator entanglement entropy for the same setups.
    From light to dark colors: $L=50,100,150,200$.
    }
    \phantomsection\label{sm_fig:cccc}
\end{figure}

\section{Different local interactions at the impurity}
\label{sec:SM_other_impurities}

In this section we show that the operator dynamics discussed in the main text is robust to the microscopic form of the impurity interaction.

In addition to the two-body density interaction
\bea
    H_{\mathrm{imp}}=\Delta N_i N_{i+1},
\eea
we also consider the following representative local impurity terms:
\bea
    H_{\mathrm{imp}}=\Delta\, N_i N_{i+1} N_{i+2},
\eea
and
\bea
H_{\mathrm{imp}} = \Delta\left( c_i^\dagger c_{i+1}^\dagger c_{i+2}^\dagger c_{i+3} + c_{i+3}^\dagger c_{i+2} c_{i+1} c_i \right).
\eea
In all cases, we study the Heisenberg evolution of the initial local operator
\bea
    O(0)=N_i,
\eea
and compute the impurity-sensitive weight $w(t)$ defined in Eq.~\eqref{eq:def_pt}, together with the half-system operator entanglement entropy.

Representative results are shown in Figs.~\ref{sm_fig:nnn} and \ref{sm_fig:cccc}. Numerically, we find that these different impurity interactions lead to the same qualitative behavior as the two-body model discussed in the main text. In particular, in the weak-coupling regime the impurity-sensitive weight continues to exhibit a power-law decay,
\bea
    w(t)\sim t^{-\alpha},
\eea
with
\bea
    \alpha=
    \begin{cases}
3, & \text{boundary impurity},\\ 1, & \text{bulk impurity}.
    \end{cases}
\eea
At the same time, the half-system operator entanglement entropy remains consistent with the upper-bound estimate from the configuration entropy. For boundary impurities, the entropy remains approximately independent of system size. For bulk impurities, the entropy instead exhibits a slow growth over the accessible time window.

\begin{figure}
    \centering
    \includegraphics[width=0.99\linewidth]{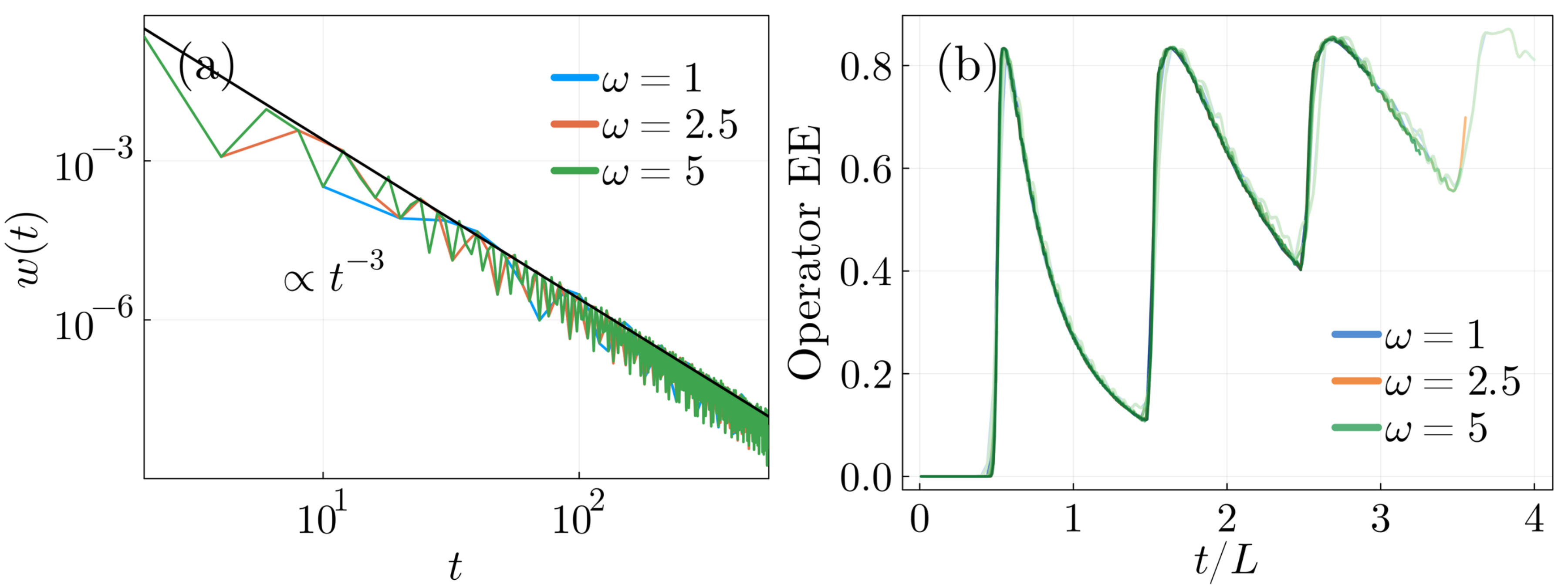}
    \caption{
    Operator dynamics in the Floquet realization of the operator model with a boundary impurity.
    Data are shown for $\omega=1,2.5,5$ at fixed impurity strength $\Delta=0.1$.
    (a) Impurity-sensitive weight $w(t)$ for $L=400$.
    The solid line indicates the power-law decay $t^{-3}$.
    (b) Half-system operator entanglement entropy.
    From light to dark colors: $L=50,100,150,200$.
    }
    \phantomsection\label{sm_fig:floquet}
\end{figure}

\section{Floquet realization of the operator model}
\label{sec:SM_floquet}
In this section we discuss a Floquet realization of the operator model as an additional robustness check of the branching picture.

The stroboscopic dynamics is generated by alternating unitary evolution under the free bulk Hamiltonian $H_0$ and the impurity interaction $H_{\mathrm{imp}}$:
\bea
    U_F=e^{-iH_0/\omega}e^{-iH_{\mathrm{imp}}/\omega},
\eea
where $\omega$ is the Floquet frequency. One Floquet period therefore corresponds to a total evolution time
\bea
    T_F=\frac{2}{\omega}.
\eea
We study the Heisenberg evolution of the initial local operator
\bea
    O(0)=N_i,
\eea
under the stroboscopic map
\bea
    O(nT_F)=\left(U_F^\dagger\right)^n O(0)\, U_F^n.
\eea

As in the static operator model, we compute the impurity-sensitive weight $w(t)$ and the half-system operator entanglement entropy. Representative results for a boundary impurity are shown in Fig.~\ref{sm_fig:floquet}. Numerically, we find that the Floquet realization exhibits the same qualitative behavior as the static Hamiltonian dynamics. In particular, for weak impurity strength the weight $w(t)$ continues to decay as a power law with the same exponent as in the static boundary case, namely $\alpha=3$, while the half-system operator entanglement entropy collapses when plotted against $t/L$, indicating that it remains approximately independent of system size over the accessible time window.

Within the range of Floquet frequencies studied numerically, the dynamics is only weakly dependent on $\omega$. This indicates that the branching picture developed in the main text does not rely on strictly static evolution, but remains valid more generally in this stroboscopic realization as well.

\begin{figure}
    \centering
    \includegraphics[width=0.99\linewidth]{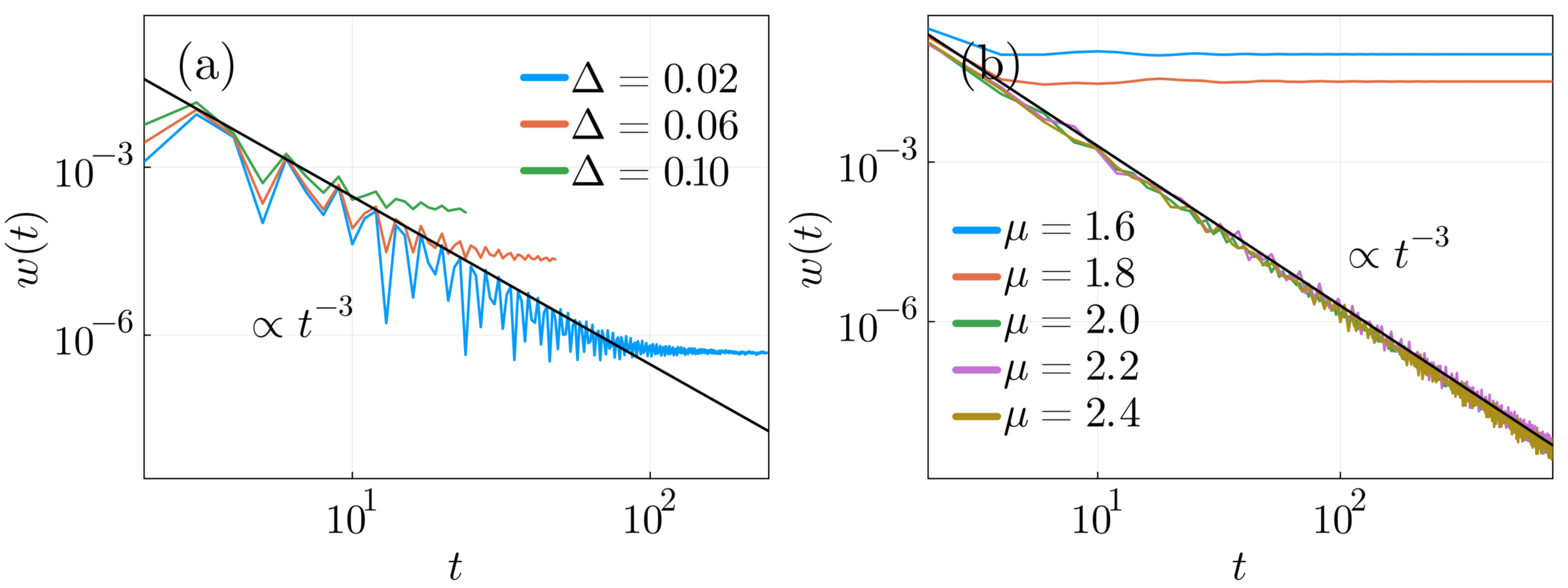}
    \caption{
    Breakdown of the effective picture.
    (a) Parity-breaking impurity at the boundary, with
    $H_{\mathrm{imp}}=\Delta(c_i^\dagger c_{i+1}^\dagger c_{i+2}+ \mathrm{h.c.})$.
    Data are shown for $L=400$.
    The impurity-sensitive weight $w(t)$ no longer follows the power-law decay of the parity-preserving case, but instead approaches a nonzero constant at long times.
    (b) Boundary-localized modes in a Kitaev-type bulk Hamiltonian, shown for the initial boundary operator $O(0)=c_1^\dagger$ and different values of $\mu$ at fixed $\lambda=1$.
    Data are shown for $L=800$.
    The topological regime ($\mu=1.6,1.8$) exhibits qualitatively different long-time behavior due to boundary modes.
    Solid lines indicate the reference decay $t^{-3}$ expected in the simple hopping model without these breakdown mechanisms.
    Results in (b) are obtained from exact free-fermion calculations.
    }
    \phantomsection\label{sm_fig:break}
\end{figure}

\section{Parity-breaking impurity}
\label{sec:SM_parity_breaking}
In this section we discuss a case where the branching picture breaks down. The parity-preserving impurities considered in the main text have an important simplification: once operator weight leaves the impurity region, it evolves under the free bulk Hamiltonian, and further impurity-induced growth can occur only when impurity-sensitive weight returns.

This locality structure fails if the impurity interaction breaks fermion parity on the impurity region. As an example, we consider
\bea
H_{\mathrm{imp}} = \Delta\left( c_i^\dagger c_{i+1}^\dagger c_{i+2} + c_{i+2}^\dagger c_{i+1} c_i \right).
\eea
In the spin representation, fermionic ladder operators carry nonlocal Jordan--Wigner strings. For a parity-breaking impurity, the impurity can act on components of the form $\eta\cdots\eta\,\mathcal O_{\rm far}$ even when the non-parity part $\mathcal O_{\rm far}$ has propagated away from the impurity region. Thus the dynamics is no longer controlled solely by the local return of operator weight to the impurity.

Consistent with this expectation, the impurity-sensitive operator weight no longer follows the power-law decay found for parity-preserving impurities. Instead, as shown in Fig.~\ref{sm_fig:break}(a), it approaches a nonzero constant.

\section{Effect of boundary-localized modes on boundary operator dynamics}
\label{sec:SM_boundary_modes}

In this section we explain how boundary-localized modes modify the long-time dynamics of boundary operators and lead to a breakdown of the free-return picture.

We consider a Kitaev-type quadratic bulk Hamiltonian,
\bea
H_0 = \sum_{j=1}^{L-1} \left(c_j^\dagger c_{j+1}+c_{j+1}^\dagger c_j\right) +\mu\sum_{j=1}^{L}N_j +\lambda\sum_{j=1}^{L-1} \left(c_j^\dagger c_{j+1}^\dagger+c_{j+1}c_j\right),
\eea
defined on a finite chain with open boundary conditions. For $\lambda\neq0$ and $|\mu|<2$, this model is in the topological phase and supports Majorana zero modes localized near the boundaries~\cite{kitaev2001unpaired}.

To understand the effect of such modes on operator dynamics, it is useful to consider the spectral decomposition of a Heisenberg operator,
\bea
O(t)=\sum_{m,n} e^{i(E_m-E_n)t}\, \ket{m}\bra{m}O(0)\ket{n}\bra{n},
\eea
where $\{\ket{m}\}$ are many-body eigenstates of $H_0$ with energies $\{E_m\}$. The long-time behavior of $O(t)$ is controlled by contributions with small energy differences $|E_m-E_n|$, and in particular by terms with $E_m=E_n$, which do not oscillate in time.

In the simple hopping model, the zero-frequency contributions from the diagonal terms in the spectral decomposition, $m=n$, can only give local operators near the impurity that are diagonal in the local fermion basis, such as $\mathbb I$ and the local parity $\eta$. Although such terms can remain near the impurity, they are not counted by the impurity-sensitive weight $w(t)$, which measures the ladder-type local operator components that can be further modified by the impurity interaction. Therefore the long-time behavior of $w(t)$ is controlled not by these diagonal pieces, but by the probability for the nontrivial operator components to return to the impurity under the free quantum walk. This gives the algebraic decay found in the main text.

The situation changes qualitatively in the presence of a boundary Majorana zero mode $\gamma$. In this case, there exist degenerate many-body eigenstates $\ket{m}\neq\ket{n}$ with $E_m=E_n$ that are connected by $\gamma$, so that
\bea
    \bra{m}\gamma\ket{n}\neq 0.
\eea
A boundary operator $O(0)$ generally has a finite overlap with $\gamma$, and therefore its time evolution contains nonoscillatory off-diagonal contributions associated with the zero mode. These contributions retain ladder-sector weight at the boundary and prevent $w(t)$ from decaying to zero at long times.

This explains the breakdown of the power-law decay found in the simple hopping model. As shown in Fig.~\ref{sm_fig:break}(b), the impurity-sensitive weight no longer decays to zero. The reason is that a boundary-localized mode traps a finite fraction of the operator weight near the impurity, so the dynamics is no longer controlled only by return under the free bulk quantum walk. Operators initialized deep in the bulk have only exponentially small overlap with the boundary zero mode and are therefore much less affected.

\end{document}